\PassOptionsToPackage{table,xcdraw}{xcolor}

\documentclass[screen,authorversion,nonacm]{acmart}

\setcopyright{rightsretained}
\acmJournal{PACMHCI}
\acmYear{2024} \acmVolume{8} \acmNumber{CSCW1} \acmArticle{41} \acmMonth{4} \acmPrice{}\acmDOI{10.1145/3637318}

\usepackage{fancyhdr}
\usepackage{multirow}
\usepackage[table,xcdraw]{xcolor}
\usepackage{color,soul}
\newcommand{\new}[1]{\textcolor{black}{#1}}

\AtBeginDocument{%
    \addtolength{\footskip}{2.0\baselineskip}%
    \fancyfoot[L]{\textit{\textbf{This is the author's version of the work. It is posted here for your personal use. Not for redistribution. The definitive Version of Record was published in {PACMHCI}, http://doi.org/10.1145/3637318.}}}%
}

\def\eg{\emph{e.g., }} 
\def\ie{\emph{i.e., }}  

\begin{document}

\title[Designing for Appropriate Reliance]{Designing for Appropriate Reliance: The Roles of \\AI Uncertainty Presentation, Initial User Decision, and \\User Demographics in AI-Assisted Decision-Making}

\author{Shiye Cao}
\affiliation{%
  \institution{Johns Hopkins University}
  \streetaddress{3400 North Charles Street}
  \city{Baltimore}
  \state{MD}
  \country{USA}
  \postcode{21218}
}
\author{Anqi Liu}
\affiliation{%
  \institution{Johns Hopkins University}
  \streetaddress{3400 North Charles Street}
  \city{Baltimore}
  \state{MD}
  \country{USA}
  \postcode{21218}
}
\author{Chien-Ming Huang}
\affiliation{%
  \institution{Johns Hopkins University}
  \streetaddress{3400 North Charles Street}
  \city{Baltimore}
  \state{MD}
  \country{USA}
  \postcode{21218}
}

\begin{abstract}
Appropriate reliance is critical to achieving synergistic human-AI collaboration. For instance, when users over-rely on AI assistance, their human-AI team performance is bounded by the model's capability. This work studies how the presentation of model uncertainty may steer users' decision-making toward fostering appropriate reliance. Our results demonstrate that showing the calibrated model uncertainty alone is inadequate. Rather, calibrating model uncertainty and presenting it in a frequency format allow users to adjust their reliance accordingly and help reduce the effect of confirmation bias on their decisions. Furthermore, the critical nature of our skin cancer screening task skews participants' judgment, causing their reliance to vary depending on their initial decision. Additionally, step-wise multiple regression analyses revealed how user demographics such as age and familiarity with probability and statistics influence human-AI collaborative decision-making. We discuss the potential for model uncertainty presentation, initial user decision, and user demographics to be incorporated in designing personalized AI aids for appropriate reliance.
\end{abstract}



\makeatother

\maketitle

\section{Introduction}
Synergistic human-AI collaboration necessitates appropriate user reliance on AI assistance. The full potential of the human-AI partnership can only be realized if the user relies on the AI when it is dependable but maintains agency in times when the AI performance may be inadequate. 
However, prior studies have demonstrated that in experimental tasks such as text classification \cite{bansal2021explanations}, deception detection \cite{lai2019human, lai2020chicago}, and treatment selection \cite{jacobs2021machine}, there is a lack of understanding among users regarding the appropriate situation and extent to rely on AI suggestions \cite{buccinca2021trust, mohseni2020prevent, karli2023if}.
Consequently, while the human-AI teams typically surpass the performance of the human working alone, they tend to fall behind the capabilities of the AI working by itself. This sub-optimal performance can have serious consequences, especially as AI systems continue to be developed and adopted to help people with making critical decisions. As an example, Google recently released their ``AI-powered'' dermatology assist tool---\textit{DermAssist}---to allow everyday users to perform at-home skin checks to identify potential skin conditions (\eg skin cancers, skin infections, acne etc.) based on photographs submitted through the tool \cite{liu2020deep, bui_liu_2021}. In critical tasks like this, it is imperative to prevent users from over-relying or under-relying on AI assistance. The underlying challenge of inappropriate reliance lies in users' difficulty in forming an accurate mental model of AI systems' capabilities, performance levels, and inner workings \cite{passioverreliance}. As AI systems are, and will continue to be, imperfect, it is paramount to design interventions aimed at enhancing user understanding of AI to help users adjust their reliance on AI assistance more appropriately \cite{cai2019hello, lai2019human}.

One common approach for helping users better understand the reliability of each AI prediction is the presentation of model uncertainty information along with its predictions. Instance-based model confidence offers users an indication of the AI's uncertainty and can be represented by metrics like the probability of the predicted label (\ie softmax output \cite{hendrycks2016baseline} or calibrated softmax output \cite{guo2017calibration,kull2017beta}). This information can help users assess the accuracy and reliability of the AI recommendation and make more informed decisions about how much to rely on the AI assistance in that task instance. \new{In particular, calibrated model softmax outputs match its actual performance on similar inputs, which makes it a more accurate representation of the reliability of the AI suggestion than uncalibrated model softmax outputs.} However, previous work in human-AI interaction has found mixed results on the effectiveness of presenting the model confidence in modulating users' reliance behavior \cite{zhang2020confidence, rechkemmer2022confidence, bussone2015role}. The effectiveness of model confidence presentation hinges on the user's ability to properly interpret the provided statistics and adjust their reliance on the AI assistance accordingly \cite{zhang2020confidence,buccinca2021trust}. Unfortunately, prior research uncovered a multitude of cognitive biases that hinders human processing of statistical information \cite{tversky1985framing, kahneman2011thinking, hamm1994underweighting, bar1980base, cossette2014heuristics, saposnik2016cognitive, stacy1995cognitive, tribe1970trial, neal2014cognitive}. Novice and expert humans alike suffer from \textit{collective statistical illiteracy}---``widespread inability to understand the meaning of numbers''--- even in critical domains such as healthcare \cite{gigerenzer2007helping}. Therefore, in this work, we ask the question: ``\textit{How may we design model uncertainty presentation to foster appropriate reliance in human-AI collaboration?}''

To answer this question, we conducted an online experiment to investigate how different presentations of the model uncertainty information (no model uncertainty information presented, raw probability-based model confidence presented, calibrated probability-based model confidence presented, calibrated frequency-based model confidence presented) may shape user reliance behavior in human-AI collaboration. 
We contextualized our experiment in AI-assisted decision-making in healthcare for laypeople, focusing on image-based skin cancer screening (similar to the idea of Google's \textit{DermAssist}); skin cancer is the most common cancer in the U.S. and monthly self-checks for skin cancer is recommended. 
In our investigation of model uncertainty presentation, we observed that, amongst participants whose initial response mismatched the AI prediction, there existed large individual differences in their likelihood of switching to agree with the AI prediction. Thus, we further explored the potential influence of \textit{AI uncertainty presentation},  \textit{initial user decision}, including initial response and its corresponding confidence, and \textit{user demographics}, such as age, gender, and familiarity with probability and statistics, on people's willingness to switch to agree with the AI suggestion given that the user and the AI initially diverged in opinions.

Our study reveals new empirical knowledge of 1) we observed no significant benefits in user reliance behavior of showing the calibrated uncertainty as opposed to the uncalibrated model confidence alone; having a calibrated model allows for the derivation of calibrated frequency presentation, which may help users more appropriately adjust their reliance and reduce confirmation bias in decision-making; 2) users have a tendency to make type 1 errors over type 2 errors in critical tasks such as cancer screening, which may significantly influence user reliance patterns; and 3) user demographics including age and user familiarity with statistics influence user reliance patterns; specifically users who self-report to be more familiar with probability and statistics may have a higher likelihood of switching to agree with AI suggestions.
Our findings suggest the need for model uncertainty presentation, initial user decision, and user demographics to be considered holistically in fostering appropriate user reliance on AI. 
Our findings further point to design implications for personalized, adaptive AI aids for critical decision support for laypeople.

\section{Related Work}
\subsection{Appropriate Reliance in Human-AI Teaming}
AI agents' performance on select critical tasks from numerous domains, \eg recidivism prediction \cite{kleinberg2018human} and cancer diagnosis \cite{wang2016deep, wang2021brilliant, bussone2015role}, are surpassing that of human experts. Thus, while ethical and legal concerns remain, AI agents are called to be used to assist humans in critical decisions \cite{cai2019hello, lai2019human}. However, users are prone to over-relying on AI assistance, hindering the human-AI team performance \cite{buccinca2021trust, mohseni2020prevent, gaube2021ai}, which can have severe consequences in critical decisions. Part of the cause of over-reliance behavior in users can be attributed to cognitive biases in decision-making \cite{lu2021reliance, pop2015individual, logg2019algorithm}; for instance, automation bias leads users to weigh AI suggestions and explanations more than information from non-automated sources particularly in objective and unfamiliar tasks (\eg weight estimation and business forecasting) \cite{logg2019algorithm}. Another major contributing factor of over-reliance on AI is the users' lack of awareness of the AI system's capabilities, performance, and inner workings \cite{passioverreliance}. 

Prior work in human-AI interaction adopted a variety of metrics to evaluate the appropriateness of reliance in users from measures of reliance. One common metric of reliance is ``human-AI agreement'', which considers the degree, number of times, or frequency at which the user's response agrees with the AI \cite{cao2023pressure, yin2019understanding, sharifheravi2020disaster, lu2021reliance, mohseni2020prevent, liu2021outofdistribution, mahmood2022effects}. In interaction paradigms where the user provides an initial decision before receiving the AI suggestion, the ``switch'' metric is commonly used. Switch captures the number of times or frequency at which an individual switched their response such that their final response matched the AI suggestion, given that the user's initial response did not match the AI suggestion \cite{cao2023pressure, yin2019understanding, lu2021reliance, merritt2011affective}. Other metrics of reliance, such as weight of advice \cite{poursabzi2021manipulating, logg2019algorithm}, how fast an AI suggestion is accepted by the user \cite{feng2019can}, user's self-reported level of reliance on AI \cite{cao2022reliance, chandrasekaran2018explanations}, and the relative length of user's gaze duration on the AI suggestion \cite{cao2022reliance}, have also been used to estimate user reliance on AI assistance. Building on these metrics, studies evaluate appropriate reliance by identifying over-reliance and under-reliance, \ie requesting automation when the reliability of the AI is lower than that of the user (over-reliance) \cite{okamura2020adaptive}, agreeing with incorrect AI suggestions (over-reliance) \cite{sutherland2015role, buccinca2021trust, bussone2015role, yang2020visual}, disagreeing with correct AI suggestions (under-reliance) \cite{sutherland2015role, yang2020visual, bussone2015role}. Inspired by prior work, in this study, we use the switch to AI suggestion to assess user reliance on AI assistance. In addition to user response, we also ask for user confidence in their initial response and final response. This allows us to evaluate AI suggestions' influence on user confidence in their responses in addition to whether or not the user switches to agree with AI suggestions. By calculating the change in user confidence between their initial response and final response, we can obtain a more nuanced measure of model influence on user or user reliance on AI assistance. Furthermore, since users in our experiment provide an initial response before being presented with the AI suggestion, we consider switching to incorrect recommendations to determine whether users are over-relying on AI assistance.

\new{\subsection{Designing for Appropriate Reliance}}
Prior works have investigated interventions aimed at reducing inappropriate reliance behavior in users during human-AI collaboration. Cognitive forcing function, such as delaying the presentation of the AI suggestion to the user or slowing down the users' decision-making process, was found to significantly reduce user over-reliance on incorrect AI recommendations \cite{buccinca2021trust, rastogi2020deciding}. Other interventions tried to help users better understand the AI's capabilities, performance, and inner workings by providing more information about the model itself or its predictions. For instance, studies explored if providing explanations can help users better understand how the AI came up with the AI suggestion, which may allow users to better assess the reliability of the AI suggestion. However, presenting explanations of the AI decisions did not appear to reduce inappropriate reliance in its users \cite{lai2019human, bansal2021explanations,liu2021understanding, poursabzi2021manipulating}. On the contrary, some studies found that explanations may increase user reliance on incorrect AI recommendations \cite{bansal2021explanations, zhang2020confidence, liu2021understanding, bussone2015role}. Moreover, prior work discovered mixed effects in presenting model uncertainty information to users on user reliance on AI suggestions. Previous studies found that showing high uncertainty in model input features can lead to an undesirably high decrease in user confidence in their decisions and their trust in the system \cite{wang2021show, lim2011investigating}. On the contrary, another study found that limited impact of presenting model confidence (written as a percentage) on user reliance in the AI, when the model confidence is presented with the model's stated accuracy on held-out data \cite{rechkemmer2022confidence}. Another study received similar results in that manipulating the level of model confidence score (below $30\%$ or above $70\%$) shown to the user had little effect on their reliance level on the AI \cite{bussone2015role}. However, previous work also showed that presenting the model confidence in frequency format helped users calibrate their trust in the AI, but did not help improve decision accuracy  \cite{zhang2020confidence}. Thus, while presenting model uncertainty information should theoretically help users estimate and assess the reliability of the AI suggestion, it is unclear from prior work how model uncertainty information affects user reliance.

\new{One possible reason for the mixed findings is that humans struggle with interpreting and acting on numbers. Cognitive biases have been shown to cause difficulty in probability inference for individuals across expertise levels \cite{hamm1994underweighting, gigerenzer2007helping}. When asked to draw conclusions about a person's health from health statistics, patients, journalists, and physicians alike showed evidence of ``collective statistical illiteracy'' without noticing \cite{gigerenzer2007helping}. Prior work in AI-assisted decision-making has also found this effect to impact the user's ability to interpret and act on the model accuracy information presented \cite{lai2019human}. The study found that any presentation of AI accuracy increases human reliance on the AI, even if the presented claimed accuracy is as low as random chance ($50\%$ accuracy in binary decision-making task). To help make statistics easier to interpret and more intuitive for human readers, previous research recommends framing statistics in frequency form rather than probability form \cite{gigerenzer1995improve, cosmides1996humans}. In fact, prior work showed that the use of frequency representations of statistics could mitigate or even invert certain cognitive biases, including over-confidence bias, conjunction fallacy, and base-rate neglect \cite{gigerenzer1991make}.} 

\new{We speculate that user interpretation of model uncertainty information may also be impacted by their ability to interpret and act on numbers. Prior work supports this hypothesis as studies found that users desire the AI uncertainty information for AI suggestions \cite{gaube2021ai, gaertig2018people}, but often find the presented information difficult to understand \cite{passioverreliance, bussone2015role}. Hence, there is a need to investigate different ways of representing the model uncertainty information. To our knowledge, no guidelines exist on how to best convey model uncertainty to the user. Towards filling this knowledge gap, in this work, we compare the differential effects of three ways to present the model confidence---1) raw AI confidence shown as a probability score; 2) calibrated AI confidence shown as a probability score; and 3) calibrated AI confidence contextualized as a frequency event---on user reliance.}

\subsection{AI Uncertainty Quantification and Calibration}
Uncertainty quantification is challenging for modern machine learning. It is well known that machine learning algorithms tend to fail when the test distribution deviates from the training distribution \cite{hendrycks2016baseline}. Worse, the outputs of a deep neural network model tend to be over-confident, causing the model outputs to display high confidence even when they are inaccurate \cite{staahl2020evaluation, goodfellow2014explaining, hendrycks2016baseline}. An important metric to evaluate the confidence generated by the model is the level of confidence calibration \cite{guo2017calibration}. 
Formally, a model $f(x) = s$ is ``perfectly calibrated on a dataset if for each of its output scores $s$, the proportion of positives within instances with model output
score $s$ is equal to $s$'' \cite{kull2017beta}. \new{In other words, if a model is perfectly calibrated, then the model's output should match its actual performance, which allows users to interpret the frequency of the predicted event actually occurring; for example, if a perfectly calibrated model predicts an event with probability of 0.9, then we should expect that event to occur in roughly nine out of ten cases similar to this case.} 

Several measurements can be used to evaluate the calibration of a model, which include reliability diagrams \cite{degroot1983comparison, niculescu2005predicting, guo2017calibration}, expected calibration error (ECE) \cite{naeini2015obtaining, guo2017calibration}, and maximum calibration error (MCE) \cite{naeini2015obtaining, guo2017calibration}. 
To help improve the calibration of model outputs, we can use various post-hoc calibration techniques (\eg temperature scaling \cite{guo2017calibration}, beta calibration \cite{kull2017beta}, isotonic regression \cite{zadrozny2002transforming}) to learn a calibration map from uncalibrated model predictions to calibrated predictions that better match the actual probability of the event on the hold-out validation data. Post-hoc calibration techniques are prevalent as they are easy to train and can work with any trained neural network structure \cite{tomani2021post}. Aside from post-hoc techniques, there is also rich literature on the training of intrinsically uncertainty-aware neural networks, which are based on sampling or intensive retraining of the model and hence less computationally efficient \cite{gal2016dropout, lakshminarayanan2017simple, dietterich2000ensemble}.
Calibrating the models is especially helpful in AI-assisted decision-making, where a model confidence score can be used to communicate the model uncertainty in its prediction to its human collaborator and helps users to identify opportunities for intervention \cite{hendrycks2016baseline}. In our experiment, we utilize beta calibration \cite{kull2017beta} (after model selection) to generate more calibrated model confidence so that we can represent the model uncertainty information in frequency form.

\section{Methods}
We designed and conducted an online user study with the \textit{model uncertainty presentation} as a between-subjects factor to understand how model uncertainty presentation affects user reliance in AI-assisted decision-making. During our analysis, we observed individual differences in participants' likelihood of switching. Therefore, we conduct additional post-hoc exploratory analysis in hopes of identifying some key factors that possibly affect the user's decision to switch during human-AI interaction. 

\subsection{Experimental Task}
We contextualized our experiment in a skin cancer screening task. Three melanoma cancer cases and 12 melanocytic nevus cases were randomly selected from the ISIC 2018 challenge dataset to serve as the experimental task \cite{codella2019skin, tschandl2018ham10000}. We picked $25\%$ to be the prevalence of skin cancer in our experiment to roughly replicate the prevalence of skin cancer in the real world \cite{guy2015prevalence, stern2010prevalence}. We chose only to include melanoma and nevus cases since the binary outcome of benign or malignant is more clinically relevant than multi-class classification since, in practice, a patient would want to get examined as soon as possible by a doctor for any type of malignancy \cite{tschandl2019comparison}. 

We chose this setting because skin cancer is the most common cancer in the U.S. and monthly at-home self-checks for skin cancer are recommended. Various AI-aided medical decision-making tools, \eg Google's DermAssist, have been developed for AI ``experts'' to assist novice users in the skin self-check and track detected skin lesions and moles for changes over time so that people can make a more informed decision about their next steps \cite{brewer2013mobile, bui_liu_2021}. AI assistants have also been designed to help primary care physicians and nurse practitioners diagnose skin conditions more accurately under impending physician shortage \cite{jain2021development}. In this study, we are interested in how novice users interact with AI assistants in the skin cancer diagnosis context and what factors influence their reliance on the AI agent. 

\subsection{AI Suggestion}
We trained a deep neural network for binary skin cancer classification to assist human users in this experiment. 

\subsubsection{Data} We randomly selected 10 melanoma and 40 melanocytic nevus cases from the ISIC 2018 challenge dataset \cite{codella2019skin, tschandl2018ham10000} for testing purposes. The three melanoma cases and 12 melanocytic nevus cases used in the main experiment were randomly drawn from this held-out test set (cancer base rate is $25\%$ in the experiment). Then, we used $60\%$ of the rest of the melanoma and melanocytic nevus cases from the ISIC dataset for model training and $40\%$ for model validation. We augmented the training data with a random horizontal flip with a probability of $.50$. 

\subsubsection{Model Training} We trained a deep neural network for binary skin cancer classification by fine-tuning a PyTorch TorchVision Resnet-18 model pre-trained on ImageNet (ImageNet 1-crop error rates: top-1 accuracy = $.70$ and top-5 accuracy = $.89$). The trained model had an accuracy of $.97$ on the training set, an accuracy of $.83$ on the validation set, and an accuracy of $.80$ on the 15 test cases used in the experiment: correctly predicting 10 out of 12 benign cases and 2 out of 3 cancer cases. 

\begin{figure}[t]
    \includegraphics[width=1\textwidth]{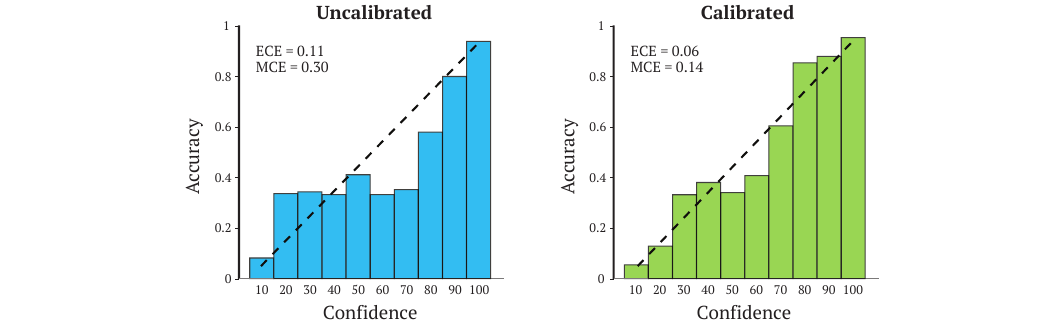}
    \caption{Reliability diagram of trained model showing a visual representation of model calibration. The diagram plots the expected sample accuracy as a function of model confidence for the cancer class on a held-out test set.}
    \Description{Reliability diagram of trained model showing a visual representation of model calibration. The diagram plots the expected sample accuracy as a function of model confidence for the cancer class on a held-out test set. The actual accuracy of samples from the confidence bins of the uncalibrated model is further from the perfect calibration line than the actual accuracy of samples from the confidence bins of the calibrated model.}
\label{fig:reliability-diagram}
\end{figure}

\subsubsection{Model Calibration}
We evaluated the calibration of our model output for the positive class using the expected calibration error (ECE) and maximum calibration error (MCE) metrics with 10 bins \cite{naeini2015obtaining, guo2017calibration}. On a held-out test set with 50 samples, our trained model achieved an ECE of $0.11$ and a MCE of $0.30$ (0 indicates perfect calibration in ECE and MCE). 

To improve the calibration of model output, we explored various post-hoc calibration methods, including beta calibration \cite{kull2017beta}, Platt scaling \cite{platt1999probabilistic}, temperature scaling \cite{guo2017calibration}, isotonic regression \cite{zadrozny2002transforming}, and Gaussian process \cite{wenger2020non}. By applying these methods, our model output became better aligned with the model performance. On the held-out test set, our model achieved an ECE of $0.06$ and MCE of $0.14$ after beta calibration; an ECE of $0.08$ and MCE of $0.16$ after
Platt scaling; an ECE of $0.05$ and MCE of $0.19$ after temperature scaling, an ECE of $0.10$ and MCE of $0.25$ after isotonic regression, and an ECE of $0.11$ and MCE of $0.27$ after gaussian process. To select the best-calibrated model, we considered a combination of ECE and MCE as the selection criteria. Based on these two metrics, we chose to present the model output after beta calibration as the model's calibrated confidence score. Figure \ref{fig:reliability-diagram} (reliability diagram) visualizes the calibration of the model outputs on the 50 samples from the held-out test set before and after beta calibration. 

In Appendix \ref{proof:classwise-calibrated}, we showed that if a binary classifier is calibrated for one class, then it is class-wise calibrated. Therefore, our classifier, calibrated for the positive class as shown in Figure \ref{fig:reliability-diagram}, is class-wise calibrated. \new{As mentioned in the related work, our class-wise calibrated classifier's model output for the positive/negative class is representative of the true likelihood or frequency of positive/negative instances among similar samples.} Calibration allows model output to be used to indicate model performance on specific instances (see Appendix \ref{appendix:confusion-matrix}), which we use to generate the calibrated frequency presentation \new{(more details on this in Section \ref{sec:model-uncertainty-presentation-calibration})}.

\begin{figure}[t]
    \includegraphics[width=1\textwidth]{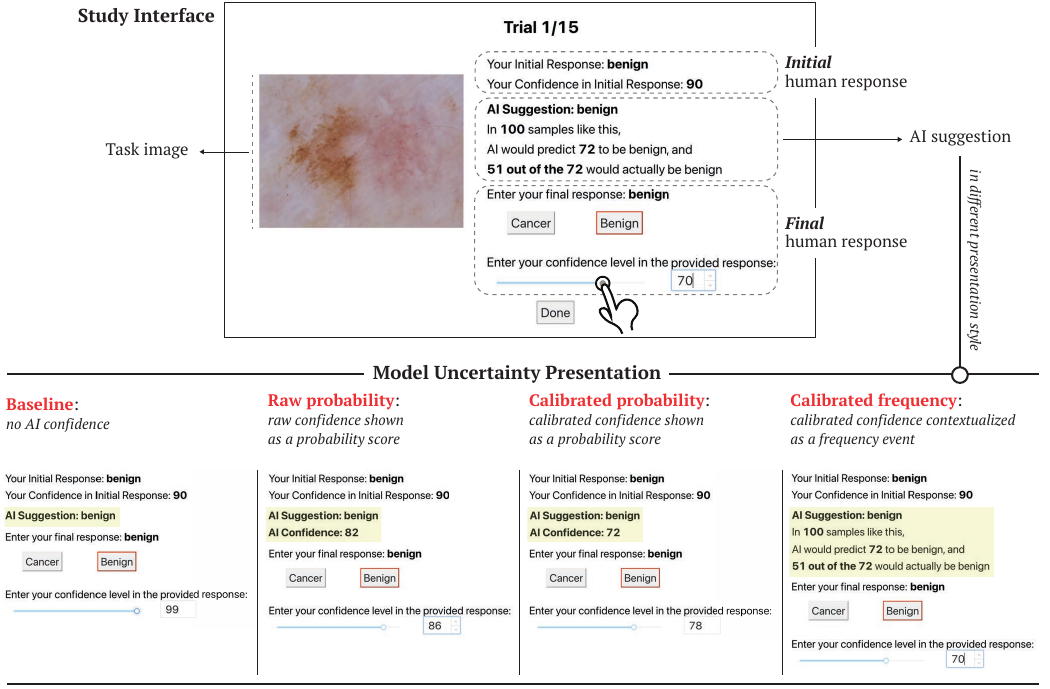}
    \caption{Overview of the four different model uncertainty presentations explored in the study.}
    \Description{Overview of the 4 different model uncertainty presentations used in the study. The figures show a screenshot of what a user would see in a trial. The task image is presented on the left, and on the right side is the user's initial response, uncertainty in their initial response, model suggestion, model uncertainty presentation, and place to enter the final response and uncertainty in the final response. In the baseline model uncertainty presentation, the model suggestion is shown with no uncertainty information. The raw probability and calibrated probability presentations are shown the same way just as a probability value. The calibrated frequency presentation presents the model uncertainty as the model performance in 100 samples like the test case ("In 100 samples like this, AI would predict 72 to be benign and 51 out of 72 would actually be benign").}
\label{fig:confidence-presentation-style}
\end{figure}

\subsection{Model Uncertainty Presentation}
\label{sec:model-uncertainty-presentation-calibration}
At the beginning of each experiment, participants were randomly assigned to one of four model uncertainty presentations (See Figure \ref{fig:confidence-presentation-style} for an example task with each of the four model uncertainty presentations): 

\begin{itemize}
    \item \textit{Baseline}: No uncertainty information presented. 
    
    \item \textit{Raw Probability}: The raw model confidence presented as a percentage (\ie AI Confidence: 82; shown second to left in Figure \ref{fig:confidence-presentation-style}). The model's average raw confidence on the 15 test cases was $0.88$ ($SD=0.10$). \new{The raw (uncalibrated) probability does not accurately match the true frequency of the predicted event given the input (or among similar samples like this).}
    
    \item \textit{Calibrated Probability}: The calibrated model confidence presented as a percentage (\ie AI Confidence: 72; shown second to right in Figure \ref{fig:confidence-presentation-style}). Since the model probability is calibrated, the value is roughly representative of the true likelihood of the prediction being correct. The model's average calibrated confidence on the 15 test cases was $0.80$ ($SD=0.12$). 
    
    \item \textit{Calibrated Frequency}: The calibrated model confidence presented in frequency form contextualized as the estimated model performance in 100 samples like the test case (\ie In 100 samples like this, AI would predict 72 to be benign, and 51 out of the 72 would actually be benign; shown right-most in Figure \ref{fig:confidence-presentation-style}). Using the calibrated model, among the cases that the model predicted to be of a specific class (number of samples $\times$ model calibrated confidence), the number of cases that would actually be of the predicted class is the number of samples $\times$ model calibrated confidence $\times$ model calibrated confidence that the calibrated model confidence value of the test instance falls into \new{(see details of this derivation in Appendix \ref{appendix:confusion-matrix})}. \new{The calibrated frequency presentation delivers the same model uncertainty information that is used in other presentations in the frequency format.} By framing the model confidence in frequency form as the model's estimated performance, we hoped that the model uncertainty would be easier for the users to interpret \new{and, in turn, induce more appropriate reliance in users}.
\end{itemize}

\new{Calibration is necessary to align the model's confidence scores with the actual frequencies of the predicted event among similar samples, enabling the derivation of the calibrated frequency model uncertainty presentations. It is important to note that since the model output from our original deep neural network model before beta calibration was poorly calibrated (See Figure \ref{result:confidence-calibration} left), it is not representative of the true frequency of the predicted event among similar samples. Therefore, we cannot derive a raw frequency model uncertainty presentation and did not have this condition in our study.}

\begin{figure}[t]
    \includegraphics[width=1\textwidth]{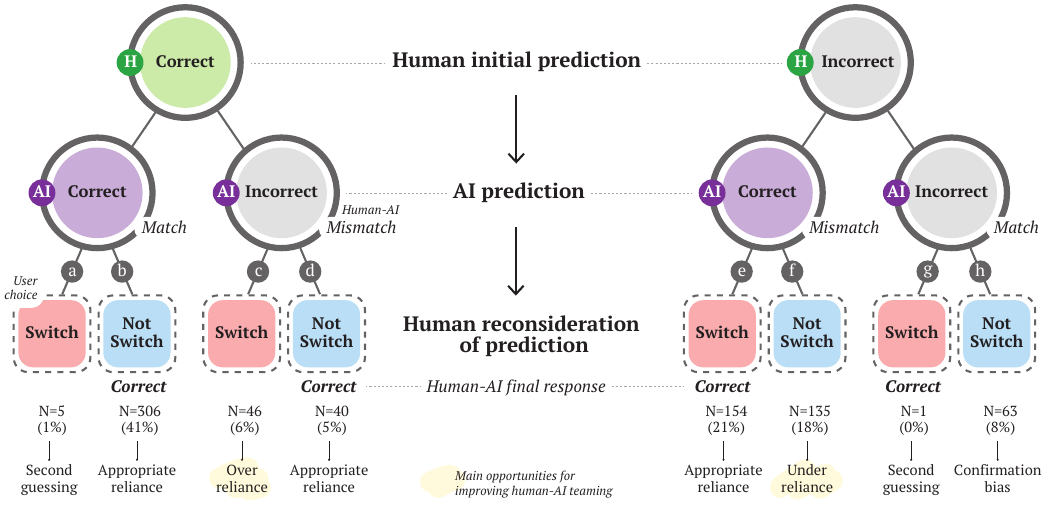}
    \caption{\new{Overview of the participants' decision-making process. Participants first provide an initial response. Then, the AI prediction is revealed to the participants. Regardless of whether the AI suggestion matched the participant's initial response, participants are asked to provide a final response on behalf of the human-AI team. The bottom of this figure shows the number/percentage of cases that belongs to each branch in our study. Then, based on the correctness of the initial response, AI suggestion, and final response, we specify whether the user choice made in that branch should be considered appropriate reliance.}}
    \Description{Describe!}
    \label{fig:flowchart}
\end{figure}

\subsection{Study Procedure}
To ensure the quality of the data, we included attention checkpoints adapted from Bauer et al. \cite{bauer2020review} during the experiment to help us screen out low-quality data. Participants were warned about the screening question (not eligible for the study if they had received prior training on skin cancer diagnosis) and attention checkpoints. Upon agreeing to participate in the study, the participants filled out a demographic survey. The demographic survey asked for the participant's gender, age, educational background, level of familiarity with AI, level of trust for AI, and if they have received prior training on skin cancer diagnosis. Then, the participants were provided with some basic rules for skin cancer classification \cite{stern2010prevalence, simon_2020_skin_cancer}. Participants were only allowed to move on to the main experiment if they were able to identify all five warning signs of skin cancer from a list of characteristics from memory (more details on user training in Appendix \ref{appendix:user-training}). 
Once the user passed the training phase, in each trial, they were first asked to provide an initial response to the task (benign or cancer) and their confidence in their initial response on a scale of 0--100. After users confirmed their initial decision, the AI suggestion and confidence score were shown together. The AI suggestion is correct in 12 out of 15 cases, and the order at which the cases were presented was randomized. How the uncertainty associated with the AI suggestion is displayed depended on the condition that the user was randomly assigned at the beginning of the experiment. Then, the user was asked to make a final decision and provide their confidence in their final decision on a scale of 0--100. After each trial, no feedback on the previous trial was given to the participant to reduce possible learning effects. After participants completed all 15 trials, they were asked to rate on a five-point Likert scale their agreement to the statement ``I understood the model confidence in the suggestions''. See Figure \ref{fig:flowchart} for the full user decision process of a task instance.

\subsection{Measures}
To investigate the effect of the different model uncertainty presentation (baseline, raw probability, calibrated probability, calibrated frequency) on user reliance behavior, we adopted the following metrics:

\new{Note that to simplify our writing, we use ``match/mismatch'' to refer to the AI suggestion matching/mismatching the user's initial response and we used ``switch/not switch'' to refer to whether or not users decided to change their response after viewing the AI suggestion.}

\begin{itemize}    
    \item \textit{Switch}: This metric captures whether the user updated their final response to agree with the AI suggestion, given that the AI suggestion mismatched their initial response. This measure is also commonly used in prior work when studying user reliance in human-AI collaboration (\eg \cite{passioverreliance, yin2019understanding, lu2021reliance, merritt2011affective}). 

    \item \textit{\new{Switch to} Incorrect Recommendations}: Whether or not the user \new{switched} to agree with the AI suggestion, given that the AI suggestion is incorrect and their initial response \new{mismatched} the AI suggestion. Inspired by prior work (\eg \cite{passioverreliance, buccinca2021trust}), this metric is used to measure over-reliance. 
    
    \item \textit{Confidence Change} (Final User Confidence - Initial User Confidence): The difference in user confidence before and after seeing the AI suggestion. \new{Ideally, user confidence should be high if their decision is correct and low if their decision is incorrect. Thus,} AI suggestion \new{matching} the initial user response should not wildly increase the user's confidence, particularly when the user's final decision is incorrect. Moreover, AI suggestion \new{mismatching} the initial user response should not lead to \new{unrestrained decrease in user confidence}, particularly when the user's final decision is correct. \new{We consider this metric a more nuanced measure of user reliance on AI in addition to considerations of human-AI agreement and switch.}

     \item \textit{Perceived Understanding of AI Uncertainty}: In the post-study survey, participants with the raw probability, calibrated probability, and calibrated frequency presentation reported their self-perceived understanding of the model uncertainty information. \new{We were particularly interested in whether there would be a difference in the participants’ self-reported understanding of uncertainty between the probability and frequency representations.}  
\end{itemize}

\subsection{Participants}
\label{sec:participants}
A total of 50 participants (27 female, 19 male, and 4 other) were recruited online through convenience sampling from the local community, using electronic newsletter posts and posts to student group mailing lists. As a result, the majority of the participants were relatively young ($M=26.16, SD=9.97$) and highly educated (20 completed high school, 17 have bachelor's degrees, 13 have master's degrees). Most participants indicated that they were somewhat familiar with AI ($M=3.60$ out of $5, SD=0.88$) and somewhat trusted AI technology ($M=3.30$ out of $5, SD=0.97$). None of the participants were medical professionals who have previously received training in skin cancer diagnosis. The majority of participants also self-reported to be somewhat familiar with statistics ($M=3.34$ out of $5, SD=0.69$). On average, the participants spent $19.33$ minutes ($SD=9.17$) to complete the study. The participants each received a $\$8.00$ gift card as compensation for their time. The study was approved by our institutional review board (IRB).

\section{Results}
\begin{figure}[t]
    \includegraphics[width=1\textwidth]{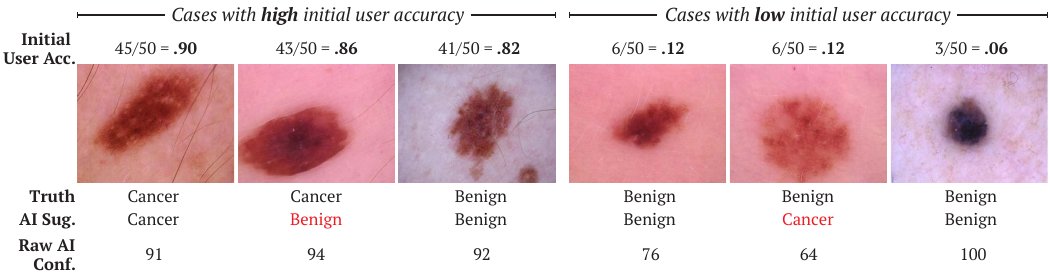}
    \caption{Example test cases from the experiment. Participants and the AI agent appeared to have complementary expertise; The AI made the incorrect prediction on a test case that most participants correctly identified as cancer, but the AI also made the correct prediction on two test cases that most participants failed to correctly classify initially.}
    \Description{Participants had high initial accuracy (0.9, 0.86, 0.82) on the three test cases to the left and low initial accuracy (0.12, 0.12, 0.06) on the three test cases to the right by themselves. The AI suggestion did not match the ground truth in two out of six cases, one of which participants had $86\%$ accuracy and the other participants had $12\%$ accuracy. This shows that participants and the AI appeared to have complementary expertise as the AI made the incorrect prediction on a test case that most participants correctly identified as cancer, but was also able to make the correct prediction on two test cases that most participants did not correctly classify initially.}
\label{fig:task-cases}
\end{figure}

\subsection{Human, AI, and Human-AI Team Performance}
\label{result:task-performance}
\new{First, we tried to understand on a high-level how well participants performed on the task by themselves compared to the AI by itself. This is important as it would impact how users use and rely on AI assistance. A better understanding of participants' performance on the task helps contextualize our understanding of appropriate reliance in this specific task context, \ie when and how much users should rely on AI assistance.}

\subsubsection{Users Err on the Side of Caution} 
\label{result:task-error-distribution}
\begin{figure}[t]
    \includegraphics[width=1\textwidth]{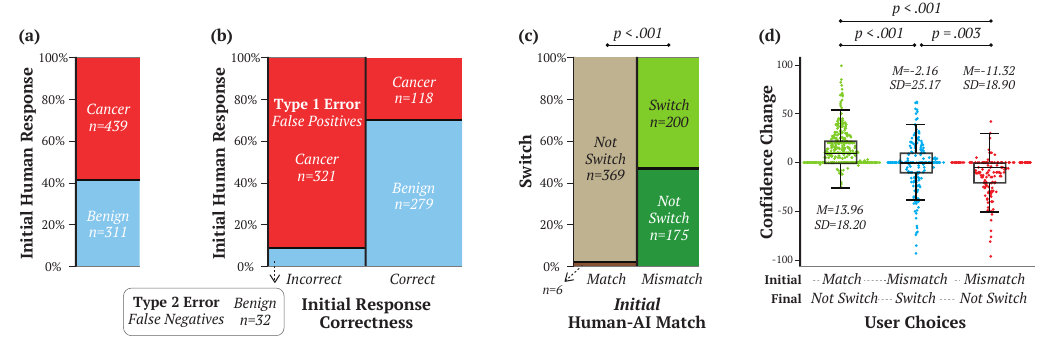}
    \caption{(a) Distribution of initial user decision. Participants were more likely to think that a case is cancer than benign, even though cancer is the less likely event. (b) Distribution of initial user decision by the correctness of the decision. Participants are much more likely to make a Type 1 error than a Type 2 error in their initial response. (c) Distribution of switch to AI by initial human-AI match. Among cases in which the AI suggestion matched the user's initial response, participants almost never switched to disagree with the AI such that their final response almost always still agreed with the AI suggestion. (d) Distribution of user confidence change among cases in which (green) the AI suggestion \new{matched} the user's initial response, and the user \new{did not switch their response to disagree with the AI suggestion (see Figure \ref{fig:flowchart} branches b, h)}; (blue) the AI suggestion \new{mismatched} the user's initial response, and \new{switched their response to agree with the AI (see Figure \ref{fig:flowchart} branches c, e)}; (red) the AI suggestion \new{mismatched} the user's initial response, and the \new{user did not switch their response to agree with the AI suggestion (see Figure \ref{fig:flowchart} d, f)}. Participants increased their confidence when the AI suggestion matched their initial response and decreased their confidence when the AI suggestion mismatched their initial response and they did not switch their response to agree with the AI. A smaller decrease in confidence was observed in cases in which the AI suggestion mismatched the user's initial response and the user switched their response to agree with the AI than cases in which the AI suggestion mismatched the user's initial response and the user did not switch their response to agree with the AI suggestion.}
    \Description{Plot a shows the distribution of initial user decisions. Participants were more likely to think that a case is cancer than benign, even though cancer is the less likely event. Plot b shows the distribution of initial user decisions by the correctness of the decision. Participants are much more likely to make a Type 1 error than a Type 2 error in their initial response. Plot c shows the distribution of the final AI agreement by the initial AI agreement. Among cases in which the AI suggestion agreed with the initial user response, participants almost always maintained their response such that their final response still agreed with the AI suggestion. Plot d shows the distribution of confidence among cases in which the green box plot shows that participants initially agreed with the AI and maintained their initial response as their final response. The blue box plot shows that participants initially disagreed with the AI and updated their final response to agree with the AI; the red box plot shows that participants initially disagreed with the AI and maintained their initial response as their final response. Participants had a significantly smaller decrease in confidence when they initially disagreed with the AI and switched than when initially disagreed with the AI and anchored in their own response.}
    \label{fig:results-basics}
\end{figure}

\new{The AI correctly predicted $80\%$ (12 out of 15 cases) in the experiment. Far below the AI performance, the participants' standalone average accuracy (average accuracy of participants' initial responses) was $53\%$ (397 out of 750 cases). The main contributing factor to poor human standalone performance is} their caution towards missing cancer. Even though participants were \new{explicitly} told that cancer is a less likely event with a base rate of $0.20$ in training \new{(see Appdenix \ref{appendix:user-training} for more details on the training users received)}, participants were more likely to provide the initial response that a case is cancer (439 out of 750 cases $= 0.59$) than benign (311 out of 750 cases $=0.41$) (Figure \ref{fig:results-basics} a). In consequence, \new{participants provided the correct response on much more cancer cases (118 out of 150 cases $=0.79$) than benign cases (279 out of 600 cases $=0.41$).} Moreover, type 1 errors (321 out of 353 initial response error cases $=0.91$; 215 out of 249 final response error cases $=0.86$) were much more prevalent than type 2 errors (32 out of 353 initial response error cases $=0.09$; 34 out of 249 final response error cases $=0.14$) among participants' initial and final response (Figure \ref{fig:results-basics} b).

\new{As AI performance on average was significantly better than the human participants, we move on to analyze how users used the AI suggestion and whether the novice humans had anything to offer to the team on the case-by-case (within-instance) level. If so, regardless of the human performance being low, the human-AI collaboration can still be synergistic under appropriate reliance, and complimentary performance would be a possibility. If not, participants should always choose to rely on AI, which limits the team performance to be strictly less than or equal to the AI-alone performance.} 

\subsubsection{Within-Instance Variance in Human Versus AI Performance}
\label{result:complementary-expertise}

\new{The average final team accuracy is $67\%$, which is better than the human-alone performance and worse than that of the AI-alone. The reason for this lack of complementary performance is that, on the case-by-case level, the team performance strictly increased from the initial human-alone performance when the AI prediction on the case was correct and strictly decreased from the initial human-alone performance when the AI prediction on the case was incorrect, demonstrating over-reliance behavior in users.}

\new{While the AI had much higher accuracy on the task, across all instances, than the participants, this difference in performance did not hold within specific instances. This is because participants had a large variance in accuracy across instances, ranging from $6\%$ to $90\%$. Notably, more than $50\%$ of the participants correctly classified two out of three cases that the AI incorrectly classified. Specifically, $86\%$ of the participants correctly classified a case (C2) with an average confidence of $.63$ that the AI misclassified as benign (raw model confidence $.94$). Moreover, $74\%$ of the participants correctly classified (average confidence $.67$) a case (C6) that the AI misclassified as cancer (raw model confidence $.82$). For the third case (C13) that the AI misclassified as benign (raw model confidence $.64$), $88\%$ of the participants also misclassified (average confidence $.67$) it. On the other hand, more than $50\%$ of participants misclassified 6 out of the 12 cases that the AI correctly classified. Among these cases, the AI provided highly confident correct predictions (raw model confidence $.96$, $1.00$) in two cases; correct but less certain predictions (raw model confidence $.76$, $.79$, $.82$) in three cases; and an incorrect prediction (raw model confidence $.64$) in one case.}

\new{Figure \ref{fig:task-cases} shows some samples of these cases. See Appendix \ref{appendix:complementarity} for a table with the human (average), AI, and team (average) performance for each task instance. These findings indicate that while no complimentary performance was observed overall, some degree of within-instance variance existed between the human and the AI performance in skin-cancer predictions, suggesting potential opportunities for leveraging human-AI collaboration on these specific task instances to achieve complimentary performance.} 

\subsection{Initial User Response and Prediction Match}
\label{result:effect-init-AI-agreement}
\new{We explored at a high-level how participants used and relied on the AI suggestion. More specifically, we sought to understand whether match\footnote{We note again that to simplify our writing, we use ``match/mismatch" to refer to the AI suggestion matching/mismatching the user's initial response and ``switch/not switch" to refer to whether or not users decided to change their response after viewing the AI suggestion.} impacted whether user switched and their confidence in their decisions. }

\subsubsection{Initial Response Match With AI Predominantly Determines Likelihood of User Switch}
\label{result:init-AI-agreement-on-final-AI-agreement}
Our data indicate that in cases where the participants' initial response matched the AI prediction, participants almost never switched (Figure \ref{fig:results-basics} c). In fact, participants only \new{switched} in 6 out of 375 cases ($2\%$) in which their initial response matched the AI suggestion. We present more details on these six cases in Appendix \ref{appendix:case-study}. Due to the rarity of participants deciding to switch when the AI suggestion matched their initial response, for the rest of our analysis, we only focused on participants whose initial response mismatched the AI suggestion and those whose initial response matched the AI suggestion and did not switch. Furthermore, we also separated our analysis on confidence change based on whether the participants' initial response matched the AI prediction in the rest of the analysis.

\subsubsection{Initial Response Match and User Decision to Switch to AI Regulate Confidence Change}
\label{result:init-AI-agreement-on-change-in-confidence}
We explored the effect of user decision choices (match and not switch, see Figure \ref{fig:flowchart} branches b, h; mismatch and switch, see Figure \ref{fig:flowchart} branches c, e; mismatch and not switch, see Figure \ref{fig:flowchart} branches d, f) on change in user confidence before and after seeing the AI suggestion. We performed a two-way repeated measure analysis of variance (ANOVA) test where decision choices were set as a within-subjects factor, uncertainty presentation as a between-subjects factor, and participants as a random effect. We discovered a significant main effect of user decision choice on confidence change, $F(2, 97.27)=34.24, p<.001$ (Figure \ref{fig:results-basics} d). Pairwise comparisons using Tukey's HSD test revealed that cases in which participants' initial response \new{matched} the AI suggestion and \new{did not switch their response to disagree} with the AI suggestion ($M=13.96, SD=18.20$) had a significantly higher confidence increase than cases in which participants \new{mismatched and switched to agree with the AI suggestion} ($M=-2.16, SD=25.17$), $p <.001$. Moreover, cases in which participants \new{matched and did not switch to disagree with the AI suggestion} had a significantly higher confidence change than those in which participants \new{mismatched and did not switch} ($M=-11.32, SD=18.90$), $p<.001$.  In addition, cases in which participants \new{mismatched and switched} had a significantly higher confidence change than cases in which participants \new{mismatched and did not switch}, $p=.003$, indicating that participants not only adjust their confidence based on whether their initial response \new{matched} the AI suggestion but also whether or not they \new{switch to agree/disagree} with the AI suggestion. No significant main effect of uncertainty presentation on confidence change ($F(3, 40.88)=1.59, p=.207$) nor interaction effect between uncertainty presentation and decision choice ($F(6, 97.14)=0.84, p=.542$) was found. 

\subsection{Effect of Model Uncertainty Presentation} 
\label{result:effect-AI-uncertainty-presentation}
\new{We now report the effects of different model uncertainty presentations on different characterizations of user reliance on AI, including \new{switch}, \new{switch to incorrect recommendation}, and confidence change. In addition, we also studied the effect of uncertainty presentation on participants' perceived understanding of AI uncertainty.} 

\begin{figure}[t]
    \includegraphics[width=1\textwidth]{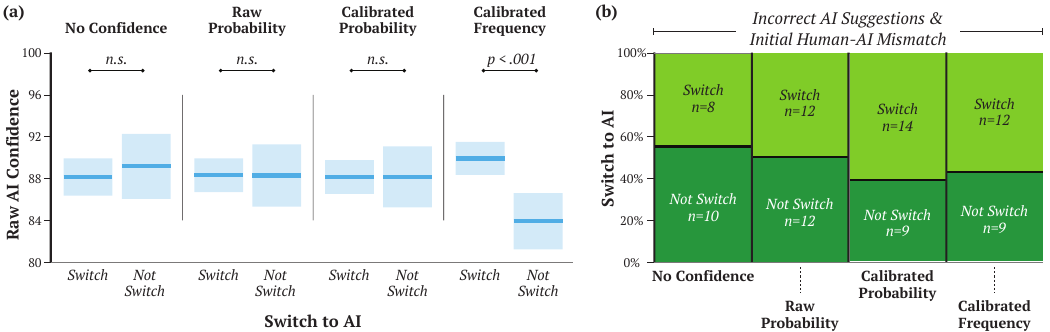}
    \caption{(a) Relationship between raw model uncertainty and whether participants decided to switch to agree with the AI suggestion (given that their initial response \new{mismatched} the AI suggestion) under different model uncertainty presentations. Participants are more likely to switch to the AI suggestion when the model confidence is higher under only the calibrated frequency presentation. (b) Given that the AI suggestion is wrong and the participants' initial response \new{mismatched} the AI suggestion \new{(see Figure \ref{fig:flowchart} branches c, d)}, this plot shows whether or not participants ultimately decided to switch to agree with the AI suggestion for each of the four model uncertainty presentations.}
    \Description{Plot a on the left shows no significant relationship between raw model uncertainty and whether participants decided to switch to agree with the AI suggestion under the no confidence, raw probability, and calibrated probability presentations. Participants are more likely to switch to agree with the AI suggestion when the AI confidence is higher under only the calibrated frequency AI uncertainty presentation. Plot b on the right shows that given that the AI suggestion is wrong and the participant initially disagreed with the wrong AI suggestion, this plot shows whether participants ultimately decided to update their initial decision to agree with AI or maintain their initial decision of disagreeing with AI, for each of the four model uncertainty presentation conditions. No significant effect of AI uncertainty presentation on reducing over-reliance was observed.}
    \label{fig:evaluating-confidence}
\end{figure}

\subsubsection{Calibrated Frequency Presentation Helps Users Regulate Their Reliance Based on Model Uncertainty}
\label{result:understanding-uncertainty}
To assess how well participants understood the model uncertainty information presented, we conducted a correlation analysis exploring the relationship between raw model confidence and the cases of switching to AI. We chose to use raw model confidence in this analysis for a fair comparison across presentation types; the calibrated model confidence and frequency-based calibrated model confidence are both derived from the raw model confidence and are alternative presentations of the uncertainty information. If participants understood the uncertainty information, we would observe that the higher the raw model confidence, the more likely it is for the participants \new{to switch to agree with the AI suggestion}. 

To run the analysis, we divided the cases in which the AI suggestion \new{mismatched} the user's initial response based on the model uncertainty presentation and fitted four logistic regression models with raw model confidence as the input variable and switch to AI as the response variable. No significant correlations between raw model confidence and switch to AI were observed in cases in which participants had the no confidence presentation, the raw probability presentation, and the calibrated probability presentation (no confidence: $\chi^2(1, 76)=0.25, p=.616$; raw probability: $\chi^2(1, 91)=0.00, p=.968$; calibrated probability: $\chi^2(1, 105)=0.07, p=.796$), suggesting that model uncertainty information did not effectively influence user reliance with these three uncertainty presentations. However, there existed a positive correlation between raw model confidence and switch under the calibrated frequency presentation, $\chi^2(1, 103)=26.05, p<.001$ (Figure \ref{fig:evaluating-confidence}, a), suggesting that participants with the calibrated frequency presentation more appropriately adjusted their reliance on the AI based on the model uncertainty information. 

\subsubsection{Model Uncertainty Presentation Does Not Reduce Over-Reliance on AI in Users}
\label{result:over-reliance}
To explore if model uncertainty presentation has an effect on over-reliance in users, we considered whether users switched among cases in which the AI suggestion was incorrect and mismatched participants' initial response \new{(Figure \ref{fig:flowchart} branches c, d)}. Contingency analysis and likelihood ratio test revealed no significant difference between the participants' likelihood to switch to the AI suggestion across the four model confidence presentations (no confidence: 8 out of 18 $=0.44$; raw probability: 12 out of 24 $=0.50$; calibrated probability: 14 out of 23 $=0.61$; calibrated frequency: 12 out of 21 $=0.57$), $\chi^2(3, 86)=1.33, p=.722$ (Figure \ref{fig:evaluating-confidence}, b). \new{In summary, over-reliance on AI was observed across all four presentation styles.}

\begin{figure}[t]
    \includegraphics[width=1\textwidth]{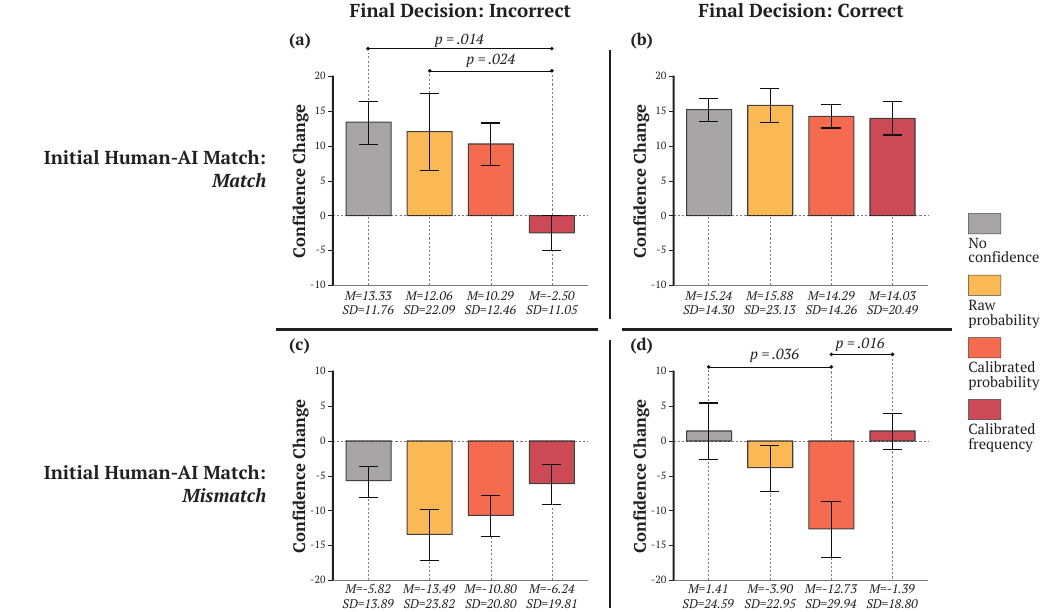}
    \caption{Effect of model uncertainty presentation on confidence change broken down by initial human-AI agreement and the correctness of the final response. Ideally, providing the model uncertainty information should help users increase their confidence if their final decision is (likely) correct and decrease their confidence if their final decision is (likely) incorrect. Our results showed that confidence changes tended to be more moderate in cases with calibrated frequency presentations than in other model uncertainty presentations. Moreover, the calibrated frequency presentation helped participants calibrate their confidence to be closer to matching the correctness of their decisions when (1) the AI suggestion \new{matches} the user's initial user response and the final response is incorrect \new{(see Figure \ref{fig:flowchart} branch h)} and (2) the AI suggestion \new{mismatched} the user's initial response and the final response is correct \new{(see Figure \ref{fig:flowchart} branches d, e)}. The error bars shown in the plots represent standard error and only significant results are emphasized.}
    \Description{This plot shows the effect of model uncertainty presentation on confidence change broken down by initial Human-AI agreement and the correctness of the final response. Confidence changes tended to be more moderate in cases with calibrated frequency presentations than in other uncertainty presentations. Moreover, the calibrated frequency presentation helped participants adjust their confidence to be closer to matching the correctness of their decisions (more appropriate) when (1) the AI suggestion matches the user's initial response and the final response is incorrect and (2) the AI suggestion mismatches the user's initial response and the final response is correct. There is no significant difference across confidence presentations when (1) the AI agrees with the initial user response and the final response is correct and (2) the AI disagrees with the initial user response and the final response is incorrect.}
    \label{fig:understanding-confidence-acc}
\end{figure}

\subsubsection{Calibrated Frequency Helps Users Adjust Their Confidence More Appropriately}
\label{result:confidence-calibration}
To explore how the model uncertainty presentations may \new{have influenced how participants adjust their confidence in their decisions}, we analyzed the participants' change in confidence with respect to whether their initial response \new{matched} the AI suggestion and whether their final response was correct (Figure \ref{fig:understanding-confidence-acc}). Ideally, we want the collaboration to work such that we would observe the following effects on confidence change: user confidence should increase when the AI suggestion \new{matches} their initial response; user confidence should decrease when the AI suggestion \new{mismatches} with their initial response; user confidence should decrease when their final decision is incorrect; and user confidence should increase when their final decision is correct. Through four one-way ANOVA tests---one for each of the four initial responses and AI matched/mismatched by the correctness of the final response conditions \new{(Figure \ref{fig:flowchart} branches b, c, d, e, f, h) to see which cases are covered by the conditions}---we studied the effect of model uncertainty presentation on user confidence change. We applied the Bonferroni correction and considered $p<.012$, which is less than $0.5$ divided by $4$ as a significant effect.

Firstly, we look at cases in which the participants' initial response \new{matched} the AI suggestion. There was a significant difference in user confidence score among cases in which the participants' initial response \new{matched} the AI suggestion but had an incorrect final response \new{(Figure \ref{fig:flowchart} branch h)} at the $p<.012$ level, $F(3, 68)=4.51, p=.006$ (Figure \ref{fig:understanding-confidence-acc} a). Pairwise post-hoc comparisons using Tukey's HSD revealed a significantly higher confidence change in cases where the participants had no AI confidence presentation ($M=13.33, SD=11.76$) than those with the calibrated frequency presentation ($M=-2.50, SD=11.05$), $p=.014$. Moreover, the confidence change was significantly higher in cases with the raw probability presentation ($M=10.29, SD=12.46$) than those with the calibrated frequency presentation, $p=.024$. Among participants whose initial response \new{matched} the AI suggestion and whose final response is correct \new{(Figure \ref{fig:flowchart} branch b)}, no significant difference between the confidence change was found across cases with varying model confidence presentations, $F(3, 307)=0.17, p=.919$ (Figure \ref{fig:understanding-confidence-acc} b). 

In addition, we look at cases in which the participants' initial response \new{mismatched} the AI suggestion. Among these examples, examples that had an incorrect final response \new{(Figure \ref{fig:flowchart} branches c, f)} show no significant difference in confidence change, $F(3, 181)=1.48, p=.220$ (Figure \ref{fig:understanding-confidence-acc} c). Among cases in which users' initial response \new{mismatched} with the AI suggestion and had a correct final response \new{(Figure \ref{fig:flowchart} branch d, e)}, there was a significant difference in confidence change, $F(3, 194)=3.81, p=.011$ (Figure \ref{fig:understanding-confidence-acc} d). Pairwise post-hoc comparisons using Tukey's HSD revealed a significantly higher confidence change in participants who had the no confidence presentation ($M=1.41, SD=24.59$) than the calibrated probability presentation ($M=-12.73, SD=29.94$), $p=.036$. Moreover, confidence change was significantly higher in cases with calibrated frequency presentation ($M=1.39, SD=18.80$) than those with the calibrated probability presentation, $p=.016$. 

\new{This result shows that the calibrated frequency presentation appeared to help participants adjust their confidence in their final response more appropriately than the other confidence presentations 1) when their initial response matched the AI suggestion and the final decision was incorrect (Figure \ref{fig:flowchart} branch h) and 2) when their initial response mismatched the AI suggestion and the final decision was correct (Figure \ref{fig:flowchart} branches d, e).}

\subsubsection{Model Uncertainty Presentation Does Not Improve Self-Reported User Understanding}
\new{We asked participants with the raw probability presentation, the calibrated probability presentation, and calibrated frequency presentation to report how confidence presentation impacted their opinions of how well they understood the model uncertainty.} Participants with the calibrated frequency presentation did not feel that they had a better understanding of the AI uncertainty than participants with the other probability presentations. Through a contingency analysis and likelihood ratio test, we did not find any significant difference between the participants' self-reported understanding of the model uncertainty across presentation styles (raw probability: $M=3.31$ out of five, $SD=0.75$; calibrated probability: $M=3.77, SD=0.93$; calibrated frequency: $M=3.54, SD=1.20$), $\chi^2(8, 39)=9.96, p=.268$. In other words, we did not find evidence supporting that users found model uncertainty information presented in frequency form to be necessarily easier for them to understand, even though users appeared to adjust their reliance behavior more properly compared to the other presentation styles.

\section{Factors that Influence User Reliance}
\label{result:exploration-of-factors}
\new{During our analysis, we observed that there were large individual differences in participants' likelihood of switching given that their initial response \new{mismatched} the AI prediction ($M=0.53, SD=0.28, min=0.0, max=1.0$).} To explore what factors influenced \new{users' decision to switch to agree with the AI suggestion} during the interaction, we used a stepwise multiple regression approach that prior works have used to understand how a range of factors shapes the task outcome \cite{huang2014multivariate, soroush2012hybrid, brendgen1999effects, gillath2021attachment}. We excluded cases in which the user's initial response \new{matched} the AI suggestion as mentioned in Section \ref{result:effect-init-AI-agreement}. We chose backward elimination as the step-wise method \cite{huang2014multivariate}; log-likelihood ratio statistic \cite{wang2007determination} as the selection criterion; and p < .25 as the stop criterion \cite{wang2007determination, sauppe2014social}. We dummy-encoded the three nominal variables (initial response[cancer], model confidence presentation type[raw probability, calibrated probability, calibrated frequency], and gender[male, female]) for the analysis. The model had nine input variables: user initial response related variables (including (1) initial response and (2) initial confidence), AI suggestion related variables (including (3) raw AI confidence and (4) AI uncertainty presentation), and user demographic variables (including (5) age, (6) gender, (7) self-reported familiarity with statistics, (8) self-reported familiarity with AI, and (9) self-reported trust in AI). \new{We confirmed through computing the GVIM value using the vim function from the car package in R that multicollinearity was not an issue in our regression analysis since $(GVIM^{1/(2DF)})^2 < 1.27$ for all nine input factors considered \cite{fox1992generalized, buteikis2020practical}.} We did not include AI suggestion \new{as a potential factor} in the model because the AI suggestion is encoded in the user's initial response since our model only included cases where the participants' initial responses mismatched the AI.  

\begin{table}[]
\caption{Stepwise multiple logistic regression on whether or not the user will switch to agree with the AI suggestion, given that the user's initial response disagreed with the AI suggestion. We included user id as a random effect in each logistic regression model to account for repeated measures. We used backward elimination as the step-wise method, log-likelihood ratio statistic as the selection criterion and $p<.25$ as the stop criterion. (***), (**), and (*) denote p<.001, p<.01, and p<.05 with z being the z-value in logistic regression.}
\Description{Table shows results from stepwise multiple logistic regression on whether or not a user will switch to the AI suggestion given that the AI suggests mismatches the user's initial response. We used backward elimination as the step-wise method, log-likelihood ratio statistic as the selection criterion and p < .25 as the stop criterion. Five variables significantly influenced whether or not the user switched to the AI suggestion (at the $<0.05$ level): (1) raw model confidence (odds ratio: 0.96, $\chi^2(1, 374)=11.61$, $p<.001$); (2) calibrated probability presentation (odds ratio: 0.34, $\chi^2(1, 374)=9.20$, $p=.003$); (3) initial response (odds ratio: 2.12, $\chi^2(1, 374)=6.03$, $p=.016$) (4) initial confidence (odds ratio: 1.04, $\chi^2(1, 374)=31.66$, $p<.001$); (5) age (odds ratio: 0.95, $\chi^2(1, 374)=17.26$, $p<.001$); (6) familiarity with probability and statistics (odds ratio: 0.48, $\chi^2(1, 374)=15.11$, $p<.001$).}
\begin{tabular}{l|ll|ll|l}
\textbf{Class}                                                                                       & \multicolumn{2}{l|}{\textbf{Predictor}}                                                                                                 & \textbf{\begin{tabular}[c]{@{}l@{}}Odds\\ Ratio\end{tabular}} & \textbf{\begin{tabular}[c]{@{}l@{}}Confidence\\ Interval\\ (95\%)\end{tabular}} & \textbf{p} \\ \midrule[1.5pt]
\multirow{4}{*}{\textbf{\begin{tabular}[c]{@{}l@{}}Model\\ Uncertainty\\ Presentation\end{tabular}}} & \multicolumn{2}{l|}{\textbf{Model Confidence}} & 1.05 & 1.02–1.08 & \textless{}.001\mbox{***} \\ \cline{2-6} & \multirow{3}{*}{\textbf{\begin{tabular}[c]{@{}l@{}}Model \\ Uncertainty\\ Presentation\end{tabular}}} & {[}raw prob.{]} & 
2.90 & 0.87–9.74 & .084 \\ & & 
\textbf{{[}calibrated prob.{]}} & 4.20 & 1.23–14.25  & .021\mbox{*} \\ &                                                     & {[}calibrated freq.{]} & 2.47 & 0.74–8.31 & .143 \\ \midrule[0.7pt] \multirow{2}{*}{\textbf{\begin{tabular}[c]{@{}l@{}}Initial User\\ Decision\end{tabular}}} & 
\multicolumn{2}{l|}{\textbf{Initial Response}} & 0.40  & 0.21–0.78 & .007\mbox{***} \\ \cline{2-6} 
& \multicolumn{2}{l|}{\textbf{Initial Confidence}} & 0.96 & 0.94–0.97 & \textless{}.001\mbox{***} \\ \midrule[0.7pt]
\multirow{2}{*}{\textbf{\begin{tabular}[c]{@{}l@{}}User\\ Demographics\end{tabular}}} & \multicolumn{2}{l|}{\textbf{Age}}    & 1.06 & 1.01–1.10 & .011\mbox{**} \\ \cline{2-6} & \multicolumn{2}{l|}{\textbf{\begin{tabular}[c]{@{}l@{}}Familiarity with Probability \\ and Statistics\end{tabular}}} & 2.27 & 1.22–4.23 & .010\mbox{**}          
\end{tabular}
\label{table:MLR-switch-disagree}
\end{table}

We used whether users switched to agree with the AI suggestion in their final response as the output variable ($200/375=53.3\%$ switched and $175/375=46.7\%$ did not switch). Three variables were removed in the following order: gender ($\chi^2(2, 374)=0.78$, $p=.676$), trust in AI ($\chi^2(1, 374)=0.22$, $p=.637$), and familiarity with AI ($\chi^2(1, 374)=0.38$, $p=.538$). The final model ($\chi^2(8, 374)=69.70$, $p<.001$) has a McFadden's Pseudo R-Square of $.142$ \footnote{McFadden's Pseudo R-Square ``tend to be considerably lower than those of the R-Square index and should not be judged by the standards for a 'good fit' in ordinary regression analysis. For example, values of 0.2 to 0.4 [for McFadden's Pseudo R-Square] represent an excellent [model] fit.'' \cite{mcfadden2021quantitative}}. Six variables significantly influenced whether or not the user switched (at the $p<.05$ level): participants were significantly more likely to switch to agree with AI suggestion when (1) the model confidence was higher ($\chi^2(1, 374)=14.83$, $p<.001$); (2) they had the calibrated probability presentation (Model uncertainty presentation: $\chi^2(2, 374)=6.19$, $p=.103$); (3) their initial response was benign ($\chi^2(1, 374)=8.16$, $p=.004$); (4) their initial confidence was lower ($\chi^2(1, 374)=39.24$, $p<.001$);  (5) they were older ($\chi^2(1, 374)=7.36$, $p=.007$); and (6) their familiarity with probability and statistics was higher ($\chi^2(1, 374)=7.52$, $p=.006$). Table \ref{table:MLR-switch-disagree} shows details of the final logistic regression model. 

\section{Discussion}
\subsection{Within-Instance Complementarity between Human and AI}
To some extent, our findings (Section \ref{result:complementary-expertise}) demonstrated that there may exist within-instance complementarity between human and AI. Among the 6 out of 12 cases that less than $50\%$ of the participants correctly classified, the AI correctly classified 5 out of the 6 cases. On the other hand, the majority of participants correctly classified 2 out of 3 cases that the AI incorrectly classified (see Appendix \ref{appendix:complementarity}). \new{This finding shows that human and AI may have different ways of processing visual information that lead them to be good at different cases \cite{tikhomirov2023medical}, opening up opportunities for attaining complementary performance in human-AI team with appropriate reliance.}

Observation of within-instance complementarity was particularly surprising in this study as only novice users unfamiliar with skin cancer screening were recruited (average human-alone task accuracy was $53\%$), indicating that even ``weak humans'' may have something to offer to the human-AI team in specialized tasks. We speculate that this may be because humans with contextual knowledge (see Appendix \ref{appendix:user-training}) are better at certain cases that may involve domain shifts. Prior works have shown that deep neural networks are prone to fail under dataset shifts \cite{quinonero2008dataset, pooch2019can, hendrycks2016baseline}, and lack the contextual knowledge and commonsense reasoning that humans have \cite{lake2017building, rastogi2022unifying}. Our conjecture aligns with a prior exploration that observed amateur players-machine teaming can out-perform machines alone and grand-masters alone in chess \cite{kasparov2010chess}. Future work should investigate how the capabilities of users with varying expertise levels and deep neural networks converge and diverge in collaborative tasks. A better understanding of the strengths and weaknesses of people and AI models can inform the design of more productive human-AI teamwork.

\new{Within-instance complementarity grants the opportunity for complimentary performance if the human is able to leverage AI strength on top of their own strength. However, the human-AI team performance only surpassed the human-alone performance and not the AI-alone performance, due to over-reliance and under-reliance in participants on the AI when the AI suggestion was incorrect (Section \ref{result:complementary-expertise}). Task performance strictly increased with correct AI prediction and strictly decreased with incorrect AI prediction, which demonstrated over-reliance in participants when the AI suggestion was incorrect. Although even when the AI prediction was correct with high confidence, participants did not always switch to agree with the AI, demonstrating under-reliance. This observation showed that participants struggled to gauge when to rely on the AI prediction. As a result, the human-AI team performance was sub-optimal.} 

\new{\subsection{Users Tend to Trust AI More Than Themselves}}
Prior work found that people often use the AI suggestion as a second opinion to validate their own conception and to calibrate their confidence in their responses \cite{cao2022reliance, bansal2021explanations}. Furthermore, people were more confident in their response when they perceived the AI suggestion to be in agreement with them \cite{cao2022reliance}. In alignment with prior work, we observed that participants almost \new{never switched to disagree} with their initial response and the AI suggestion when the AI suggestion \new{matched} their initial response (Figure \ref{fig:results-basics} c); additionally, participants tended to increase their confidence when the AI \new{suggestion matched} their initial response (Figure \ref{fig:results-basics} d). Among cases at which the AI suggestion and user initial response \new{mismatched}, participants had a smaller decrease in their confidence when they switched to the AI suggestion than when they did not switch ((blue) and (red) in Figure \ref{fig:results-basics} d). This may be \new{a result of our participants being novices in the experimental task} and human's general tendency to trust advice \cite{gaube2021ai}, causing users to bestow more trust in external advice than themselves. As a result, when the user's initial response mismatched the AI suggestion, participants did not decrease their confidence as much when they switched to agree with the AI suggestion than if they did not switch.

\subsection{Benefits of Calibrated Frequency Presentation of Uncertainty}
\subsubsection{\new{Calibration alone shows limited benefits: calibrated probability v.s. uncalibrated probability}} 
We did not observe apparent benefits in using the calibrated probability presentation as opposed to the uncalibrated probability presentation in our study. Users of both the raw probability and the calibrated probability presentations consistently increased their confidence when the AI suggestion \new{matched} their initial response (Figure \ref{fig:understanding-confidence-acc} a, b) and consistently decreased their confidence when the AI suggestion \new{mismatched} their response (Figure \ref{fig:understanding-confidence-acc} c, d). 
Furthermore, there was no significant difference in the amount of user confidence change between users of the two probability presentations. More specifically, participants with the raw probability presentation did not increase/decrease their confidence more because the AI \new{matched/mismatched} their initial response with higher confidence than the participants with the calibrated probability presentation. Part of the reason for this observation may be that participants showed difficulty interpreting the model confidence information presented as a probability and did not use the information to adjust their reliance on the model prediction (Figure \ref{fig:evaluating-confidence} a). This observation is in line with findings from prior work, which found little effect of presenting model confidence in probability form on user reliance on the AI \cite{bussone2015role, rechkemmer2022confidence}. 

We speculate that another reason why calibration alone may have demonstrated limited benefits might be that the difference between the raw model confidence and the calibrated model confidence presented in the 15 test instances was relatively small ($M=8.13, SD=3.29$). User reliance behavior may not have been sensitive enough to display significant behavioral changes as a result of a small difference in model confidence (\ie for the second example from the right in Figure \ref{fig:task-cases}, the raw model confidence being 94 and calibrated model confidence being 85 may have induced the same user behaviors). A recent work, however, showed benefits of calibrating the model uncertainty presentation to human behavior on top of model performance \cite{vodrahalli2022uncalibrated}; results from the user study showed that presenting over-confident model suggestions (modified according to a user behavior model) as opposed to well-calibrated model suggestions improved the accuracy and confidence of the human’s final prediction after seeing the AI advice \cite{vodrahalli2022uncalibrated}. This prior finding may explain why simply calibrating the model confidence score in our study without considering the user perspective might have led to limited benefits. 

\subsubsection{Calibrated Frequency Helps Users Adjust Their Confidence More Appropriately}
Compared to both the raw and calibrated probability presentations, we observed benefits in presenting the model confidence as a calibrated frequency. While participants did not consider the calibrated frequency presentation to be easier to understand than the two probability presentations, people's likelihood to switch was positively correlated with the raw model confidence under the calibrated frequency presentation (Figure \ref{fig:evaluating-confidence} a). In other words, participants with the calibrated frequency presentation relied on the AI more when the AI was more confident in its prediction. This observation suggests that the calibrated frequency presentation better communicated the model confidence to the users than the raw probability and calibrated probability presentations. This finding is supported by prior work that showed statistics framed in frequency form is more intuitive for people to understand and helps improve people's statistical inferences ability \cite{gigerenzer1995improve, cosmides1996humans, gigerenzer1996psychology}. 

Moreover, our results showed the use of the calibrated frequency presentation helped alleviate the effects of confirmation bias (Figure \ref{fig:understanding-confidence-acc}). Prior work showed that under the influence of confirmation bias, the discovery of evidence in favor of one's judgment exacerbates the individual's over-confidence bias \cite{koriat1980reasons}. However, this effect was not observed in participants with the calibrated frequency presentation. While all participants clearly increased their confidence among cases at which their initial response \new{matched} the AI suggestion and their final response was correct (Figure \ref{fig:understanding-confidence-acc} b, \new{Figure \ref{fig:flowchart} branch b}), only participants with the calibrated frequency presentation did not increase their confidence among cases where the participant's initial response \new{matched} the AI suggestion but their final response was incorrect (Figure \ref{fig:understanding-confidence-acc} a, \new{Figure \ref{fig:flowchart} branch h}). This behavior showed that participants with the calibrated frequency presentation tended to be less over-confident than participants with the other presentation styles. Similarly, we found that all participants whose initial response \new{mismatched} the AI suggestion and their final response was incorrect decreased their confidence (Figure \ref{fig:understanding-confidence-acc} c, \new{Figure \ref{fig:flowchart} branches c, f}); yet, among participants whose initial response mismatched the AI suggestion but their final response was correct \new{(Figure \ref{fig:flowchart} branches d, e)}, only those with the baseline and calibrated frequency presentation did not decrease their confidence on average (Figure \ref{fig:understanding-confidence-acc} d). Together, these observations showed that the calibrated frequency presentation helped users adjust their confidence in their responses more appropriately than other uncertainty presentations. 

\new{\subsubsection{Calibrated frequency does not prevent over-reliance}}
\new{Different from observations of the positive effects of calibrated frequency presentation on user confidence change, presenting the model confidence, in any form, did not help prevent over-reliance in users---switching to} the AI suggestion when the AI is incorrect (Figure \ref{fig:evaluating-confidence} b) \cite{buccinca2020proxy}. Participants tended to over-rely and frequently \new{switched to agree with} incorrect AI suggestions ($73\%$), regardless of how the model confidence was presented. To our surprise, presenting the calibrated, usually visibly lower, model confidence (in the calibrated probability and calibrated frequency presentation) did not cause users to be less susceptible to over-relying on the AI compared to when the raw model confidence or no model uncertainty information was presented (Figure \ref{fig:evaluating-confidence} b). This may be because all the participants in this study were novices in the task. As a result, they were more prone to agreeing (569 out of 750 cases $=76\%$) and, without proper intervention, over-relying on the AI suggestion. This finding is consistent with previous research that found higher levels of user reliance on AI assistance when the users are less certain \cite{bansal2021explanations} or have lower domain expertise in the task \cite{gaube2021ai}.

\new{In summary, we observed little benefit to showing the calibrated probability presentation over the uncalibrated probability presentation. Benefits were only observed in showing the calibrated frequency presentation. Considering when users switched to agree with the AI suggestion, the calibrated frequency presentation helped users better regulate their decisions based on the model uncertainty, suggesting a better understanding of model uncertainty information. With respect to confidence change, calibrated frequency presentation helped users adjust their confidence more appropriately than the other presentations. However, calibrated frequency presentation did not demonstrate the ability to help users avoid switching to agree with incorrect AI suggestions. It is also worth mentioning that the calibrated frequency presentation did not demonstrate any negative effects on any characterizations of user reliance considered in this study.}

\subsection{Effects of Model Uncertainty Presentation, Initial User Decision, and User Demographics on Reliance}
\new{As large individual differences in participants’ tendency to switch to AI were observed, we explored what factors influenced participants’ decisions to switch.}

\subsubsection{Effect of Model Uncertainty Presentation on User Reliance}
Our results revealed a significant effect of model confidence score and model uncertainty presentation on whether or not participants switched to agree with the AI suggestion. In particular, we found that among participants whose initial response mismatched with the AI, higher model confidence was associated with a higher likelihood of switch (Table \ref{table:MLR-switch-disagree}); this finding is in agreement with results from prior work \cite{vodrahalli2022humans} using a proxy AI (based on aggregated human response from a ``prior group of human participants''). 
Previous work found that people tend to equate confidence with competence \cite{tetlock2016superforecasting} and believe that more confident advisors possess more knowledge \cite{price2004intuitive}, suggesting that participants would prefer more confident advice. Our analysis revealed that the calibrated model with a probability presentation of uncertainty increased the odds of the participants switching to the AI suggestion compared to the models in the baseline and using the raw probability and calibrated frequency presentations (Table \ref{table:MLR-switch-disagree}).  

\subsubsection{Effect of Initial User Decision on Reliance}
Our results showed that participants initially erred on the conservative end. They were much more likely to indicate that a case is cancer than benign (Figure \ref{fig:results-basics} a), even though they were told the prevalence of cancer during training is relatively low---``studies estimate that 1 in 5 Americans will develop skin cancer in their lifetime''. As a result, the likelihood of their initial response being correct tended to be low ($M=0.53, SD=0.11$), as participants made a lot more type 1 errors ($321$ out of $353$ cases $=0.91$) than type 2 errors ($32$ out of $353$ cases $=0.09$) in their initial response (Figure \ref{fig:results-basics} b). We speculate that participants were conservative because of the high-stakes nature of the cancer screening task, causing them to deem Type 2 errors (missing cancer) to have more severe consequences than Type 1 errors (miss-diagnosing cancer). This observation resembles the decision patterns of physicians: ``\textit{when in doubt, continue to suspect illness}'' \cite{scheff1963decision}. However, type 1 errors in real-world medical diagnosis can lead to ``unjustified anxiety in patients'' \cite{scheff1963decision}, ``unnecessary expenses on more medical examinations'' \cite{scheff1963decision}, and in some cases even unnecessary surgeries \cite{bakwin1945pseudodoxia}.

In our study, AI assistance helped users recognize and reduce the number of type 1 errors in their final response ($215$ out of $249$ cases $=0.86$). However, participants continued to err on the side of caution in their final response. Our result indicated that participants were significantly more likely to switch to the AI suggestion when they initially perceived the case to be benign than when they initially believed the case to be cancer (Table \ref{table:MLR-switch-disagree}). Thus, the false positive rate in the final user response remained high and the final response accuracy ($M=0.67, SD=0.13$) was lower than the AI accuracy, suggesting room for better calibration of user reliance on AI. We also found that higher initial user confidence was associated with lower odds of the participant switching to the AI suggestion (Table \ref{table:MLR-switch-disagree}). This finding is consistent with the results of prior work \cite{chong2022human}, which found that users with higher initial confidence are less likely to adopt AI suggestions.

\subsubsection{Effect of User Demographics on Reliance}
Our analysis suggested that two user demographic factors, age and self-perceived familiarity with probability and statistics, significantly influenced user's likelihood to switch to the AI suggestion. Among cases in which the AI suggestion disagreed with the initial user response, older participants and people who are more familiar with probability and statistics have \new{higher odds} of switching to agree with the AI suggestion, an indication of a higher degree of reliance on the AI (Table \ref{table:MLR-switch-disagree}).  This finding is inconsistent with prior work that found being younger is positively correlated with trust in AI agents \cite{gillath2021attachment}, which is highly correlated with reliance and adoption of AI advice \cite{buccinca2021trust, vereschak2021evaluate}. One possible explanation is that our participants were relatively young with only six participants above the age of 35, even though their ages ranged between 19 and 60. Moreover, since probability and statistics knowledge is useful for understanding the AI confidence, participants who are less familiar with probability and statistics may have trouble interpreting the AI confidence and either ignore it or misuse it, which may potentially lead to decreased odds in switching to the AI suggestion. However, note that even though our participants had a rather wide range of self-perceived familiarity with statistics, they were well-educated (Section \ref{sec:participants}). Thus, while we recommend including age and user familiarity with probability and statistics (user profile) in predictive models to gauge when and how much people would rely on AI advice, future work should further investigate the impact of age and user statistical knowledge on user reliance with a more diverse user population.

\begin{figure}[t]
    \includegraphics[width=1\textwidth]{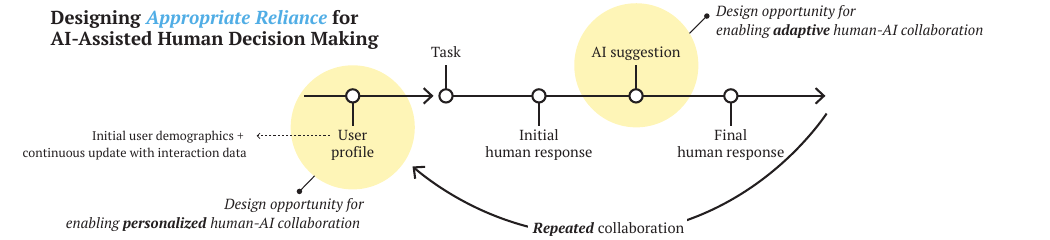}
    \caption{Overview of ideal personalized adaptive AI-Assisted human decision-making. User demographic information will be collected before the user is introduced to the AI. Collected user demographics will be to personalize the AI based on human behavior model trained on data from past users. During the task, based on the initial user response and initial user confidence, the AI will adaptively modify its presentation and what information it provides to users for optimal task behavior. Lastly, data collected and new knowledge gained about this specific user (behavioral tendencies, preferences, etc.) during the past interaction will be incorporated into the user demographics to allow for better personalizing of the AI assistance.}
    \Description{Plot shows the overview of an ideal personalized adaptive AI-Assisted human decision-making designed for appropriate reliance. User profile and demographic information will be collected before the user is introduced to the AI. Collected user demographics will be to personalize the AI based on a human behavior model trained on data from past users. During the task, based on the initial user response and initial user confidence, the AI will adaptively modify its presentation and what information it provides to users for optimal task behavior. Lastly, data collected and new knowledge gained about this specific user (behavioral tendencies, preferences, etc.) during the past interaction will be incorporated into the user demographics to allow for better personalizing of the AI assistance.}
    \label{fig:design-implication}
\end{figure}

\subsection{Towards a Design Framework for Enabling Appropriate Reliance in Human-AI Teams}
Informed by prior research and the results of this work, we here illustrate a design framework for enabling appropriate reliance in collaborative human-AI teams (Figure \ref{fig:design-implication}); the framework encapsulates the integrated use of model uncertainty presentation, initial user decision, and user demographics in designing appropriate reliance to support human-AI teamwork.
An envisioned AI model can tailor its assistance to a specific user through the consideration of relevant user demographics collected during the user onboarding stage. During the co-decision-making process, the initial user decision can be taken into account and the AI may adapt its model uncertainty presentation accordingly to ensure desirable task outcomes. The repeated collaboration process will further allow continuous updates to modules guiding user personalization and suggestion adaptation (highlighted in yellow in Figure \ref{fig:design-implication}).
To illustrate this framework, consider a 19 year-old individual who has little knowledge of probability and statistics in our study context:

\textit{Adaptation.} 
Imagine that during collaborative decision-making, this user gives an initial response (benign) and initial confidence (86) while the AI predicts that the case is cancer with high confidence (98).
The AI model recognizes that the user disagrees with its prediction with moderate confidence and predicts that given this initial user response and confidence, the likelihood of the user adopting the AI suggestion is lower than ideal. 
As a result, the model may adapt its presentation of suggestion and uncertainty as a calibrated confidence in probability form for this specific task instance. The adaptation of AI uncertainty presentation allows the AI suggestion to be given fair consideration, thereby reducing the chance of user over-relying or under-relying on the AI's advice.

\textit{Personalization.}
Given initial user demographics (younger age and little knowledge of statistics and probability), the hypothetical AI model predicts that the user would exhibit a lower level of reliance on the AI. However, after a few interactions, the AI agent might notice that this user has increased their trust in the AI.
The model may then update its setting to this specific user and decide that it would be most beneficial to present the calibrated model confidence in a frequency form rather than a probability form to reduce their chance of over-reliance on the AI.

Beyond model uncertainty presentation, there are other types of interventions that may also help users more appropriately adjust their reliance, including local \cite{zhang2020confidence, schaffer2019can, bansal2021explanations} and global \cite{wu2020towards, schrouff2021best} explanations, cognitive forcing functions \cite{buccinca2021trust}, model performance-related information \cite{rechkemmer2022confidence, yin2019understanding}, and so on. Future work should explore how interventions may be selectively used and combined to encourage appropriate reliance for a specific user in AI-assisted decision-making. Moreover, future work should investigate how we can extract user-specific information from their previous interactions with the AI agent and use it to fine-tune the user profile model and relevant model parameters.

\subsection{Limitations}
\new{This study has limitations that should be taken into account when interpreting the findings.} One limitation is the small sample size and the homogeneous population in terms of education, cultural background, and level of expertise in the experimental task. Our study involved 50 locally recruited participants. As a result, the participant population predominately consisted of young, highly educated current students who live in Western culture. As a result, we did not consider education level and cultural background in our analyses, even though these factors may influence people's task behaviors; for instance, previous research has shown that cultural background can affect people's risk-taking behaviors \cite{kreiser2010cultural}. We acknowledge that the limited sample size and homogeneity of participants may restrict the generalizability of the findings to a larger, more diverse population. Future works should explore the effectiveness of different AI uncertainty presentation styles on a wider population, \eg people who live in eastern culture, people without a college education, or older adults.

Additionally, our study was contextualized in a high-stakes skin cancer screening task involving only novices. Therefore, our findings may not generalize well to other domains, particularly those with low stakes (\eg casual games) and those that involve domain experts (\eg practicing physicians). Moreover, the benefits of the calibrated frequency presentation observed in this study are limited to short-term interactions (15 trials), which aligns with the envisioned use case of tools like Google's DermAssist. Users likely would not conduct at-home skin self-check at a frequency of more than once a month and the number of trials per use is most likely below 15. Nevertheless, future work should investigate whether our findings on the benefits of the calibrated frequency presentation would generalize to other task domains and longer interactions over time.

It is important to note that our exploratory analysis was conducted in hopes of identifying key factors that could possibly influence user reliance in AI-assisted human decision-making rather than developing an all-inclusive model for modeling user reliance. In fact, the factors discussed in this work are not exhaustive, and the coefficients in the presented model should only be interpreted in the context of our study setup and may not be applicable to other populations and tasks. However, we hope that the correlations between the identified factors and user reliance could inform future researchers in human-AI interaction of the need to consider possible influencing factors in their system design and analysis.

Moreover, as in any (online) user studies, it is difficult to ensure (even with incentives) that the participants make their best efforts in the study. In our study, we did not provide additional incentives to participants, which could have led to unintended noises in our results.

Lastly, due to limited data available for training, the AI model used in the study was not perfectly calibrated for the calibrated probability and calibrated frequency presentations. As a result, a proxy for actual frequencies of the predicted event among similar samples was used to derive the calibrated frequency presentation (see more details in Appendix \ref{appendix:calibration}).

In summary, future research should consider these limitations and further explore the generalizability of our findings to more diverse populations, different task domains and contexts, and longer interaction sessions.

\section{Conclusion}
In this paper, we present empirical findings from an online user study that explores the effect of model uncertainty presentation, initial user decision, and user demographics on user reliance on AI during assisted decision-making in a skin cancer screening task. 
Our work shows the potential benefits of representing the calibrated model confidence using the frequency form. In particular, our findings indicate that this model uncertainty presentation helps users better adjust their reliance and reduces the effect of confirmation bias on their decisions. \new{However, presenting the calibrated model confidence as opposed to the uncalibrated model confidence in probability form shows limited benefits.}
Furthermore, participants' initial decision affected their willingness to adopt the AI suggestion as the AI-assisted participants recognized and reduced their tendency to make type one errors. 
Additionally, we found that user demographics such as age and familiarity with probability and statistics influence users' reliance patterns. 
These factors have the potential to be incorporated into designing personalized AI aids for appropriate reliance.
Altogether, this work offers an empirical understanding of the role model uncertainty presentation, initial user decision, and user demographics play during AI-assisted decision-making on a high-stakes specialized task with novice users and points toward the possibility of adaptive personalized human-AI collaboration. 

\begin{acks}
This work was supported by the National Science Foundation award \#1840088 and the Malone Center for Engineering in Healthcare at the Johns Hopkins University. We would like to thank Jaimie Patterson for her feedback and assistance in this work.
\end{acks}

\newpage

\bibliographystyle{ACM-Reference-Format}
\bibliography{references}

\received{January 2023}
\received[revised]{July 2023}
\received[accepted]{December 2023}

\appendix
\section{Case Study: Human-AI Complementarity}
\label{appendix:complementarity}
\begin{table}[H]
\caption{Details on the (average) performance and (average) confidence of the human, AI, and human-AI team. The human accuracy column of cases in which the human had below-average ($<50\%$) performance on are highlighted in red. The AI accuracy column of cases in which the AI prediction was wrong on are also highlighted in red. The Team vs. Human column of cases in which the human performance surpassed that of the team are highlighted in red.}
\Description{}
\begin{tabular}{
>{\columncolor[HTML]{FFFFFF}}l 
>{\columncolor[HTML]{FFFFFF}}l 
>{\columncolor[HTML]{FFFFFF}}l 
>{\columncolor[HTML]{FFFFFF}}l 
>{\columncolor[HTML]{FFFFFF}}l 
>{\columncolor[HTML]{FFFFFF}}l }
\hline
{\color[HTML]{333333} Task} & {\color[HTML]{333333} Truth}  & {\color[HTML]{333333} \begin{tabular}[c]{@{}l@{}}Human Acc\\ (Conf.)\end{tabular}} & {\color[HTML]{333333} \begin{tabular}[c]{@{}l@{}}AI Acc\\ (Conf.)\end{tabular}} & {\color[HTML]{333333} \begin{tabular}[c]{@{}l@{}}Team Acc\\ (Conf.)\end{tabular}} & {\color[HTML]{333333} \begin{tabular}[c]{@{}l@{}}Team vs. \\ Human\end{tabular}} \\ \midrule[1.5pt]
{\color[HTML]{333333} C1}   & {\color[HTML]{333333} Cancer} & {\color[HTML]{333333} 90\%(68)}                                                    & {\color[HTML]{333333} 1(92)}                                                    & {\color[HTML]{333333} 98\%(77)}                                                   & {\color[HTML]{333333} Team}                                                      \\
{\color[HTML]{333333} C2}   & {\color[HTML]{333333} Cancer} & {\color[HTML]{333333} 86\%(63)}                                                    & {\color[HTML]{FE0000} 0(94)}                                                    & {\color[HTML]{333333} 46\%(58)}                                                   & {\color[HTML]{FE0000} Human}                                                     \\
{\color[HTML]{333333} C3}   & {\color[HTML]{333333} Benign} & {\color[HTML]{333333} 60\%(62)}                                                    & {\color[HTML]{333333} 1(100)}                                                   & {\color[HTML]{333333} 88\%(79)}                                                   & {\color[HTML]{333333} Team}                                                      \\
{\color[HTML]{333333} C4}   & {\color[HTML]{333333} Benign} & {\color[HTML]{333333} 66\%(63)}                                                    & {\color[HTML]{333333} 1(93)}                                                    & {\color[HTML]{333333} 88\%(72)}                                                   & {\color[HTML]{333333} Team}                                                      \\
{\color[HTML]{333333} C5}   & {\color[HTML]{333333} Benign} & {\color[HTML]{FE0000} 48\%(61)}                                                    & {\color[HTML]{333333} 1(82)}                                                    & {\color[HTML]{333333} 76\%(58)}                                                   & {\color[HTML]{333333} Team}                                                      \\
{\color[HTML]{333333} C6}   & {\color[HTML]{333333} Benign} & {\color[HTML]{FE0000} 12\%(67)}                                                    & {\color[HTML]{FE0000} 0(64)}                                                    & {\color[HTML]{333333} 4\%(70)}                                                    & {\color[HTML]{FE0000} Human}                                                     \\
{\color[HTML]{333333} C7}   & {\color[HTML]{333333} Benign} & {\color[HTML]{333333} 62\%(62)}                                                    & {\color[HTML]{333333} 1(99)}                                                    & {\color[HTML]{333333} 88\%(72)}                                                   & {\color[HTML]{333333} Team}                                                      \\
{\color[HTML]{333333} C8}   & {\color[HTML]{333333} Benign} & {\color[HTML]{333333} 74\%(58)}                                                    & {\color[HTML]{333333} 1(94)}                                                    & {\color[HTML]{333333} 88\%(68)}                                                   & {\color[HTML]{333333} Team}                                                      \\
{\color[HTML]{333333} C9}   & {\color[HTML]{333333} Benign} & {\color[HTML]{FE0000} 12\%(66)}                                                    & {\color[HTML]{333333} 1(76)}                                                    & {\color[HTML]{333333} 28\%(60)}                                                   & {\color[HTML]{333333} Team}                                                      \\
{\color[HTML]{333333} C10}  & {\color[HTML]{333333} Benign} & {\color[HTML]{333333} 50\%(55)}                                                    & {\color[HTML]{333333} 1(80)}                                                    & {\color[HTML]{333333} 78\%(60)}                                                   & {\color[HTML]{333333} Team}                                                      \\
{\color[HTML]{333333} C11}  & {\color[HTML]{333333} Benign} & {\color[HTML]{333333} 82\%(61)}                                                    & {\color[HTML]{333333} 1(92)}                                                    & {\color[HTML]{333333} 92\%(75)}                                                   & {\color[HTML]{333333} Team}                                                      \\
{\color[HTML]{333333} C12}  & {\color[HTML]{333333} Benign} & {\color[HTML]{FE0000} 28\%(59)}                                                    & {\color[HTML]{333333} 1(96)}                                                    & {\color[HTML]{333333} 74\%(57)}                                                   & {\color[HTML]{333333} Team}                                                      \\
{\color[HTML]{333333} C13}  & {\color[HTML]{333333} Benign} & {\color[HTML]{333333} 74\%(62)}                                                    & {\color[HTML]{FE0000} 0(82)}                                                    & {\color[HTML]{333333} 32\%(60)}                                                   & {\color[HTML]{FE0000} Human}                                                     \\
{\color[HTML]{333333} C14}  & {\color[HTML]{333333} Benign} & {\color[HTML]{FE0000} 44\%(58)}                                                    & {\color[HTML]{333333} 1(79)}                                                    & {\color[HTML]{333333} 66\%(61)}                                                   & {\color[HTML]{333333} Team}                                                      \\
{\color[HTML]{333333} C15}  & {\color[HTML]{333333} Benign} & {\color[HTML]{FE0000} 6\%(71)}                                                     & {\color[HTML]{333333} 1(100)}                                                   & {\color[HTML]{333333} 56\%(63)}                                                   & {\color[HTML]{333333} Team}                                                      \\ \hline
\end{tabular}
\end{table}

\newpage

\section{Case Study: AI Agreed with User Initially but User Changed Their Final Response to Disagree with Their Initial Response}
\label{appendix:case-study}
\begin{table}[h!]
\caption{Details of three out of six cases (see in other three cases in Table \ref{case-study-p2}) in which the AI suggestion agreed with the initial user response, but the user changed their final response to disagree with their initial response. Each of these three cases was from a different participant. Some evidence shows that these three cases may be low-quality data points. Row 4, 5, and 6 list the user's initial decision, the AI suggestion, and the final user decision for each case. After these values, the user's initial confidence, the AI's confidence, and the user's final confidence are also presented inside the parenthesis. Task time is recorded in seconds.}
\centering
\begin{tabular}{l|l|l|l}
\textbf{Case ID}                                                                                        & \textbf{Case 10}                                                                                                                       & \textbf{Case 10}                                                                                                                                                                                            & \textbf{Case 8}                                                              \\ \midrule[1.5pt]
\cellcolor[HTML]{DAE8FC}\textbf{Ground Truth}                                                           & Benign                                                                                                                             & Benign                                                                                                                                                                                                  & Benign                                                                    \\ \hline
\cellcolor[HTML]{DAE8FC}\textbf{\begin{tabular}[c]{@{}l@{}}AI Uncertainty \\ Presentation\end{tabular}} & \begin{tabular}[c]{@{}l@{}}Calibrated Frequency\end{tabular}                                                                     & \begin{tabular}[c]{@{}l@{}}Calibrated Frequency\end{tabular}                                                                                                                                          & \begin{tabular}[c]{@{}l@{}}Calibrated Frequency\end{tabular}            \\ \hline
\cellcolor[HTML]{DAE8FC}\textbf{\begin{tabular}[c]{@{}l@{}}Initial User \\ Decision\end{tabular}}       & Benign (70)                                                                                                                        & Benign (51)                                                                                                                                                                                             & Benign (50)                                                               \\ \hline
\cellcolor[HTML]{DAE8FC}\textbf{AI Suggestion}                                                          & Benign (80)                                                                                                                        & Benign (80)                                                                                                                                                                                             &                           Benign (94)          \\ \hline
\cellcolor[HTML]{DAE8FC}\textbf{\begin{tabular}[c]{@{}l@{}}Final User \\ Decision\end{tabular}}         & Cancer (70)                                                                                                                        & Cancer (51)                                                                                                                                                                                             & Cancer (20)                                                               \\ \hline
\cellcolor[HTML]{DAE8FC}\textbf{Task Time}                                                              & 98s                                                                                                                                & 12s                                                                                                                                                                                                     & 41s                                                                       \\ \hline
\cellcolor[HTML]{DAE8FC}\textbf{Notes}                                                                  & \begin{tabular}[c]{@{}l@{}} out of distribution \\ task time; \\ participant likely\\ distracted \end{tabular} & \begin{tabular}[c]{@{}l@{}} participant spent 9 seconds \\ per task in study on average\\ (less than 90\% of participants); \\ participant likely in a rush \end{tabular} & \begin{tabular}[c]{@{}l@{}}very low final \\ user confidence\end{tabular}
\end{tabular}
\label{case-study-p1}
\end{table}

\begin{table}[h!]
\caption{Details of the other three cases in which the AI agreed with the user's initial response, but the user changed their final response to disagree with their initial response. Each of these three cases was from a different participant. Some evidence shows that the first case from the left may be a low-quality data point.}
\centering
\begin{tabular}{l|l|l|l}
\textbf{Case ID}                                                                                        & \textbf{Case 14}                                                                                                                   & \textbf{Case 1}                                                                                                                                                                                                                                       & \textbf{Case 6}                                                                                                                                                                                                                                      \\ \midrule[1.5pt]
\cellcolor[HTML]{DAE8FC}\textbf{Ground Truth}                                                           & Benign                                                                                                                             & Cancer                                                                                                                                                                                                                                                & Benign                                                                                                                                                                                                                                               \\ \hline
\cellcolor[HTML]{DAE8FC}\textbf{\begin{tabular}[c]{@{}l@{}}AI Uncertainty \\ Presentation\end{tabular}} & Raw Probability                                                                                                                    & \begin{tabular}[c]{@{}l@{}}Calibrated Probability\end{tabular}                                                                                                                                                                                      & \begin{tabular}[c]{@{}l@{}}Calibrated Frequency\end{tabular}                                                                                                                                                                                       \\ \hline
\cellcolor[HTML]{DAE8FC}\textbf{\begin{tabular}[c]{@{}l@{}}Initial User \\ Decision\end{tabular}}       & Benign (80)                                                                                                                        & Cancer (60)                                                                                                                                                                                                                                           & Cancer (68)                                                                                                                                                                                                                                          \\ \hline
\cellcolor[HTML]{DAE8FC}\textbf{AI Suggestion}                                                          & Benign (79)                                                                                                                        & Cancer (81)                                                                                                                                                                                                                                           & Cancer (64)                                                                                                                                                                                                                                          \\ \hline
\cellcolor[HTML]{DAE8FC}\textbf{\begin{tabular}[c]{@{}l@{}}Final User \\ Decision\end{tabular}}         & Cancer (80)                                                                                                                        & Benign (60)                                                                                                                                                                                                                                           & Benign (42)                                                                                                                                                                                                                                          \\ \hline
\cellcolor[HTML]{DAE8FC}\textbf{Task Time}                                                              & 91s                                                                                                                                & 10s                                                                                                                                                                                                                                                   & 27s                                                                                                                                                                                                                                                  \\ \hline
\cellcolor[HTML]{DAE8FC}\textbf{Notes}                                                                  & \begin{tabular}[c]{@{}l@{}}out of \\ distribution\\ task time; \\ participant \\ likely \\ distracted \end{tabular} & \begin{tabular}[c]{@{}l@{}} low trust for AI; among seven \\ cases in which the user's initial \\ response and the AI suggestion \\ disagreed, they only switched \\ to agree with the AI in 1 out of \\ 7  cases\end{tabular} & \begin{tabular}[c]{@{}l@{}}relatively low final user \\ confidence; only case in \\ which the incorrect AI \\ suggestion matched the \\ initial response, but the \\ user updated their final \\ response to be correct\end{tabular}
\end{tabular}
\label{case-study-p2}
\end{table}

\newpage

\section{User Training}
\label{appendix:user-training}
\begin{figure}[h!]
    \includegraphics[width=1\textwidth]{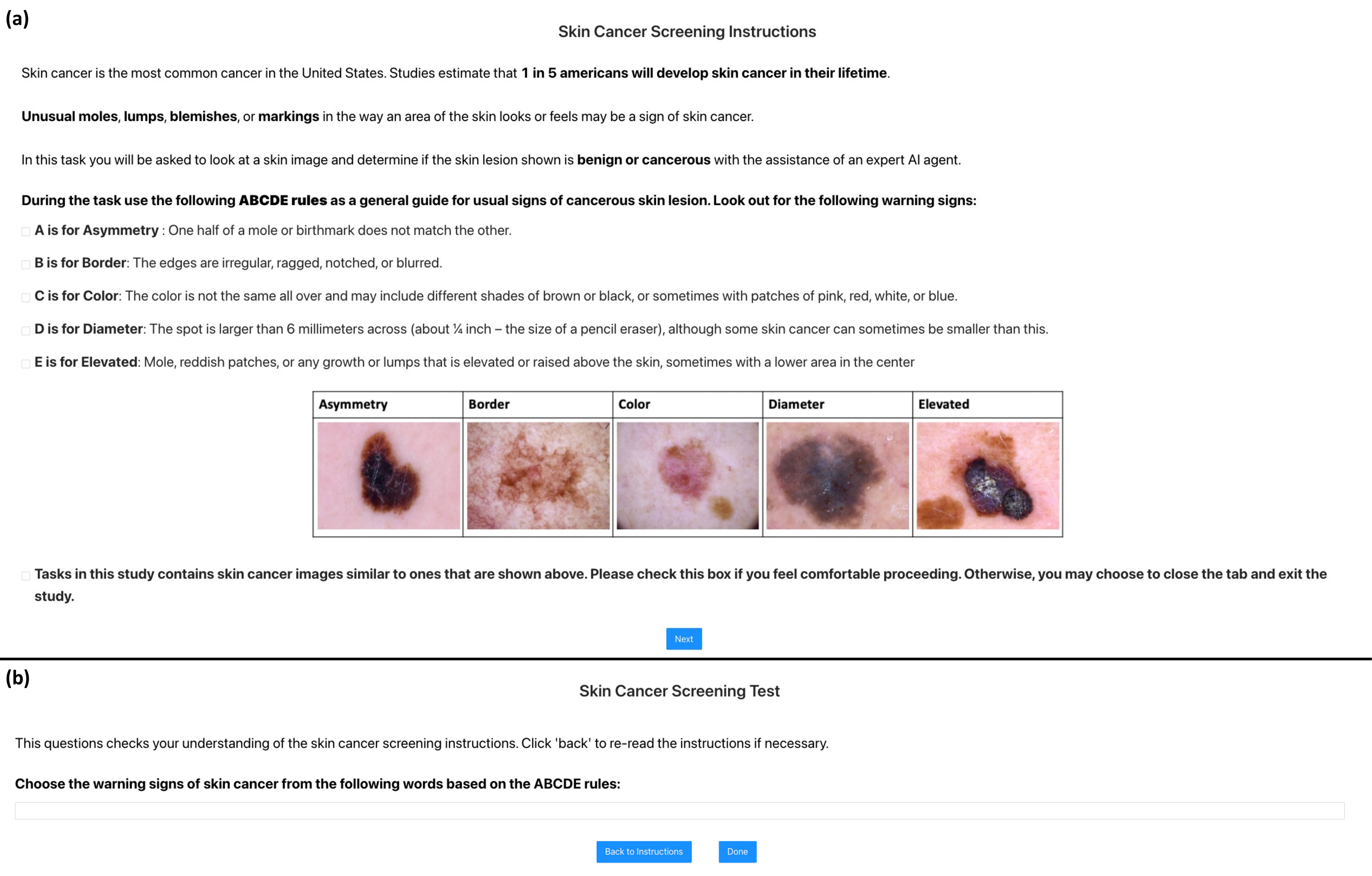}
    \caption{User skin cancer classification training: (a) Interface showing the instructions that the users were given at the start of the experiment explaining [1] the prevalence of skin cancer \cite{stern2010prevalence}; [2] task instruction; [3] warning signs of skin cancer \cite{simon_2020_skin_cancer} (we did not include the evolving aspect as a warning sign in this study because users only see one image of each case, and, therefore, cannot evaluate if the case has been evolving). To slow down the users, the users were required to click on a checkbox in front of each warning sign as they read the instructions; (b) the users were tested on whether or not the have the warning signs of skin cancer memorized. The users were only allowed to move onto the experiment if they were able to identified all five and only the five warning signs of skin cancer (asymmetry, border, color, diameter, elevated) from a list of 10 words (asymmetry, bright, border, bumpy, color, dark, diameter, elevated, hair, symmetry).}
    \Description{...}
    \label{fig:training-page}
\end{figure}

\newpage

\section{Derivation of Calibrated Frequency Model Uncertainty Representation}
\label{appendix:calibration}
\subsection{Proof: If a binary classifier is calibrated for one class, then the classifier is classwise-calibrated.}
\label{proof:classwise-calibrated}

\textbf{Proof.} Let \(\hat{p}\) be a binary \(\{0, 1\}\) probabilistic classifier that is calibrated for the positive class (1). Then, by definition \cite{kull2019beyond}, for all \( q_1 \in [0, 1] \), we have \( P[Y=1 \mid \hat{p_1}(X) = q_1] = q_1 \). Since \(\hat{p}\) is a binary classifier, let \( q_0 = 1 - q_1 \) denote the probability of the negative class.

\begin{align*}
P[Y = 0 \mid \hat{p_0}(X) = q_0] & = P[Y = 0 \mid \hat{p_1}(X) = (1 - q_0)] \\
& = 1 - P[Y = 1 \mid \hat{p_1}(X) = (1 - q_0)] \\
& = 1 - (1 - q_0) \\
& = q_0
\end{align*}

By definition, $\hat{p}$ is also calibrated for the negative class ($0$), and, thereby, classwise calibrated\cite{kull2019beyond}. Vice versa,  if a binary probabilistic classifier is calibrated for the negative class. Then, it holds for $\hat{p}$ that for any class $i$ and any predicted probability $q_i$ for this class: $P(Y = i \mid \hat{p_i}(X)=q_i) = q_i$. Thus, by definition \cite{kull2019beyond}, $\hat{p}$ is classwise calibrated. 

\subsection{Computing the Calibrated Frequency using Calibrated Models}
\label{appendix:confusion-matrix}
Let $\hat{p}$ be a binary ({0, 1}) probabilistic classifier that is calibrated for the positive class ($1$), such that $\forall q_1 \in [0, 1]$, $P[Y=1 \mid \hat{p_1}(X)=q_1] = q_1$ \cite{kull2019beyond}. Suppose we are interested in the accuracy of the model prediction on a test instance $X_{test}$ in which $\hat{p_1}(X_{test}) = q_1$.  Let $n$ be the number of samples where $\hat{p}$ outputs $q_1$ as the prediction. In other words, suppose we have $n$ samples ``similar" to $X_{test}$. Then, since  $\hat{p}$ is classwise-calibrated (proved in Appendix \ref{proof:classwise-calibrated}), we have the following confusion matrix for the $n$ examples: 

\begin{table}[H]
\caption{Confusion matrix for calibrated model $\hat{p}$ based on model confidence score with the number of true positives, true negatives, false positives, false negatives out of $n$ samples where $\hat{p}$ outputs $q_1$ as the prediction.}
\Description{Table 3: Confusion matrix for perfectly calibrated model based on model confidence score. number of true positive is equals to n times q1 squared; number of false positive is equals to n times q1 times (1 minus q1); number of false negative is equals to n times (1 minus q1) times q1; number of true negative is equals to n times (1 minus q1) squared. Total number of positive prediction is n times q1 and total number of negative predictions is n times (1 minus q1).}
\begin{tabular}{l|l|l|l}
                    & True Positive & True Negative & Total \\ \hline
Positive Prediction & $n \times {q_1}^2$ & $n\times q_1 \times (1 - q_1)$ & $n \times q_1$ \\
Negative Prediction &  $n\times (1 - q_1) \times q_1$ & $n \times {(1 - q_1)}^2$ & $n \times (1-q_1)$    
\end{tabular}
\end{table}

\end{document}